\documentclass[twocolumn]{aastex62}

\usepackage{hyperref}
\graphicspath{{./}{figures/}}

\accepted{}
\submitjournal{AJ}

\shorttitle{MUSE observations of NGC\,330}
\shortauthors{Carini et al.}

\begin{document}

\title{MUSE observations of NGC\,330 in the Small Magellanic Cloud\\ 
Helium abundance of bright main sequence stars \footnote{Based on observations made with the MUSE integral-field spectrograph operating at the Very Large Telescope (Sierro Paranal, Chile) during the commissioning of the instrument.}}

\correspondingauthor{Roberta Carini}
\email{roberta.carini@inaf.it}

\author[0000-0002-0786-7307]{R. Carini}
\affiliation{INAF - Osservatorio Astronomico di Roma \\
via Frascati 33, 00078 Monte Porzio Catone, Italy}

\author{K. Biazzo}
\affiliation{INAF - Osservatorio Astrofisico di Catania\\
via S. Sofia 78, 95123 Catania, Italy}
\affiliation{INAF -  Osservatorio Astronomico di Roma \\
via Frascati 33, 00078 Monte Porzio Catone, Italy}

\author{E. Brocato}
\affiliation{INAF -  Osservatorio Astronomico di Roma \\
via Frascati 33, 00078 Monte Porzio Catone, Italy}

\author{L. Pulone}
\affiliation{INAF -  Osservatorio Astronomico di Roma \\
via Frascati 33, 00078 Monte Porzio Catone, Italy}

\author{L. Pasquini}
\affiliation{ESO - European Southern Observatory\\
Karl Schwarzchild Str. 2, Garching b. Munchen, Germany}



\begin{abstract}

 We present observations of the most bright main sequence stars in the Small Magellanic Cloud stellar cluster NGC\,330 obtained with the integral field spectrograph MUSE@VLT.
 The use of this valuable  instrument allows us to study both photometric and spectroscopic properties of stellar populations of this young star cluster.
  
  The photometric data provide us a precise color magnitude diagram, which seems to support the presence of two stellar populations of ages of $\approx 18$ Myr and $\approx 30$ Myr assuming a metallicity of $Z~=~0.002$.

  Thanks to the spectroscopic data, we derive helium abundance of 10 main sequence stars within the effective radius $R_{\rm eff} = 20''$ of NGC\,330, thus leading to an estimation of $\epsilon(He)=10.93 \pm 0.05\,(1\sigma)$.
  
  The helium elemental abundances of stars likely belonging to the two possible stellar populations, do not show differences or dichotomy within the uncertainties. Thus, our results suggest that the two stellar populations of NGC\,330, if they exist, share similar original He abundances.
  
  If we consider stellar rotation velocity in our analysis, a coeval (30 Myr) stellar population, experiencing different values of rotation, cannot be excluded. In this case, the mean helium abundance $<\epsilon(He)>_{rot}$ obtained in our analysis is $11.00\pm0.05$ dex. 
  We also verified that possible NLTE effects cannot be identified with our analysis because of the spectral resolution and they are within our derived abundance He uncertainties.
  
  Moreover, the analysis of the He abundance as a function of the distance from the cluster center of the observed stars do not show any correlation.

\end{abstract}


\section{Introduction} \label{sec:intro}

NGC\,330 has been early recognized as one of the most populated (total mass of $\sim$ 3.6-3.8 $\times$ $10^4$ $M_{\odot}$, \cite{mackey,mclagh}) and brightest young stellar cluster ($\sim$ 30 Myr; e.g. \cite{sirianni02, keller00})of the Small Magellanic Cloud (SMC) \citep{arp,robertson}. The color-magnitude diagram (CMD) of the cluster discloses a blue main sequence and two clumps of supergiant stars, located in the red and blue part of the diagram, clearly recognizable and distinct from the main sequence stars. These features can be understood in terms of massive stars experiencing the core H-burning phase (main sequence) and the core He-burning phase (clumps). Moreover, the presence of a quite high fraction of Be stars is also well established \citep{feast}, which makes the interpretation of this cluster complicated and, at the same time, challenging (e.g., \citealt{caloi, grebel92,grebel96, keller99, keller00, martayan07a, martayan07b, tanabe13}). 
Nevertheless,  NGC\,330 remains a very attractive laboratory to improve our knowledge about the stellar evolution theory and the physics of intermediate mass stars born in a low metallicity environment such as the SMC. For instance, from an evolutionary point-of-view, the ratio between the number of blue and red supergiants represents a fair indicator of the relative time an intermediate mass star spends along the He-burning loop.
The quoted large fraction of Be stars ($\sim 60\%$), allows us  to investigate the role of rotation in intermediate mass stars \citep{grebel92, grebel96, lennon1996, mazzali96, keller98, keller99, maeder99,lennon2003}.
Moreover,  NGC\,330 represents an interesting cluster to investigate possible presence of multiple populations in a young, metal-poor environment like the SMC.
Multiple stellar populations are found in the literature in Galactic Globular Clusters (GGCs, e.g. \citealt{grat2012, piotto2010,  grat2019} and reference therein).
These latter systems are characterized by two (or more) groups of stars, one with the same chemical abundance  of halo stars, usually referred to as primordial or first  population (FG), and the other (the second generation, SG) enriched in He, N, Na, Al and poor  in O and Mg with respect the FG stars(e.g. \citealt{carretta09,carretta15,carretta18}). In particular, the enhancement of the He abundance explains the peculiarities in the CMDs of these systems, as the  MS split and the blue tails of the Horizontal Branch.
 The difference in helium between the populations can be as high as $\sim 0.1-0.2$\,dex in He mass fraction $Y$ (e.g. \citealt{hb,piotto,pasquini2011}). This is confirmed by \cite{marino2014} who provided  the first direct spectroscopic measurement of highly He-enhanced  ($Y\sim 0.34$) second generation stars in the Blue Horizontal Branch of the globular cluster NGC\,2808.
The new paradigm for the interpretation of these observations is that the GGCs are composed by two (or more) populations of stars, the  SG stars were born from matter expelled by evolving stars of the  FG stars and nuclearly processed through the hot CNO cycle.
According to the theoretical scenarios that try to explain the phenomena of the multiple populations in GCs, the process of formation of the second generation stars  happened within 100-150 Myr from the formation of the FG stars \citep{dercole2008, dercole2016, decressin, bastian2, gieles18}. 

In the recent years, the presence of multiple populations in Magellanic Clouds (MCs) Globular Clusters (GCs), similar to those found in the Milky Way, has been demonstrated by observational evidences. It was indeed found that many star clusters in the Magellanic Clouds show bimodal or extended main sequence turnoff (eMSTO) and dual clump in their color magnitude diagrams, suggesting the presence of the multiple populations of stars with possible different ages (e.g., \citealt{mackey07, glatt, girardi09, milone09, goud14}). 
More recently, in young stellar clusters (age less than 600 Myr) have been observed  not only the eMSTO,  but also the split of the MS, similar to that detected in Milky Way GCs \citep{milone15, milone2016, milone17, milone18}. These photometric evidences could be interpreted with the presence of  stellar populations with different ages,  but also with  coeval populations with different stellar rotation velocity (e.g., \citealt{bastian09, brandt15, dantona15, marino18}).
Until now, spectroscopic studies  of stars in MCs clusters younger than $\sim$ 2 Gyr do not show differences in  light elements  between stellar populations of young massive clusters \citep{mucciarelli14, martocchia17}, contrary to what is observed in the GGCs.
Only recently, \cite{lagioia} have found traces of He-enhancement ($\Delta Y \sim 0.010 \pm 0.006 $ dex) in the second populations stars of four $\sim$ 6-10 Gyr extragalactic GCs (NGC\,121, NGC\,339, and NGC\,416)  within the SMC. This has been confirmed by the analysis of \cite{chant}.
Moreover, in stars of  GCs in the MCs with age between 2 and 10 Gyr, difference in nitrogen abundance have been observed\citep{lagioia19}.

From these evidences, it is not clear if  the young massive stellar clusters  with multiple populations are the younger counterparts of the old GGCs. In this context, we started a project to investigate the stellar content and the He abundance of the stars populating the SMC young cluster NGC\,330.

The presence of multiple populations in NGC\,330 is supported by the analysis of \cite{chiosi95}, who found for the cluster a spread in age of 10-25 Myr  from classical semi-convective models, and 10-48 Myr from full and diffusive overshoot models. \cite{li2017} found that the rotation alone is not able to explain the eMSTO of NGC\,330. They also pointed out that an age spread of 35-50 Myr can help in minimizing the problem. Instead, \cite{boden} found an age between 35 and 40 Myr. So the presence and/or the origin of the age spread remains still unknown.


The Multi Unit Spectroscopic Explorer (MUSE; \citealt{bacon}) operating at the Very Large Telescope (VLT) of the European Southern Observatory (ESO)  provided us the opportunity of obtaining simultaneously photometric and spectroscopic data for a large number of stars in NGC\,330. Thanks to the exceptional observing capabilities of this instrument, in this work we investigate the presence of multiple populations and possible star-by-star differences in helium abundance.

In the present paper, we describe the observations and data reduction in Sect. \ref{sec:obs}, while the analysis of the photometric data are presented in Sect.3, together with the comparison to isochrones of different ages and rotation velocity to obtain hints on the presence of multi-populations. The spectroscopic data analysis and the measurement of He abundances are presented in Sect. 4. The effect of stellar rotation is also considered in this section. The discussion on the age of the cluster and on the radial distribution of He abundances are reported in Sect. 5, while the final remarks are provided in Sect. 6. 

\section{Observations and data reduction} \label{sec:obs}

NGC\,330 was observed with the integral-field spectrograph MUSE (\citealt{kelz}) operating at the VLT during the commissioning run in August 3rd, 2014. The observation covered the central area of NGC\,330, with a field of view of $\sim 1 \times 1$ arcmin$^{2}$, a pixel scale of 0.20 arcsec/pixel and a total exposure time of 200 s. For this observation the normal Wide Field Mode (WFM) was used and the parameters of the instrument setup adopted are summarized in Table\,\ref{musepar}. 

\begin{table}
\begin{center}
\caption{Selected parameters of MUSE in a nutshell and relevant quantities of the observations.}
\label{musepar}
\begin{tabular}{cc}
\hline
Parameter & Value \\
\hline
Number of IFU models & 24 (images slicer + \\
  & spectrograph + CCD)\\
Wavelength coverage &4800-9300 \AA\, (nominal range)\\
Field of view & 59''$\times$60'' (WFM)\\
Spatial sampling & 0.2'' (WFM)\\
Multiplex factors & 1152 slices, 90000 spaxels, \\
  & 3700 wavelength bins\\
RA (J2000) & 00:56:18.2\\
DEC (J2000) & $-$72:27:47.8\\
Mean seeing&  1.3$\arcsec$\\
Airmass& 1.5$\arcsec$\\
$T_{\rm exp}$ & 200 s \\
Mean spectral resolution &  2.75 \AA \\
\hline
\end{tabular}
\end{center}
\end{table}

\begin{figure}
\center
\includegraphics[trim= .1cm 1cm 1cm 2cm, clip=true,width=.9\columnwidth]{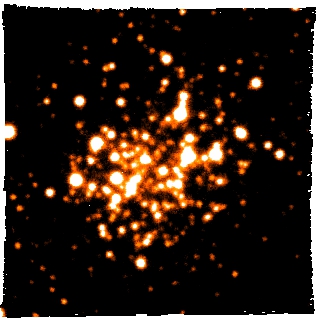}
\vspace{1cm}
\caption{MUSE image of NGC330 at $\lambda$ $\sim 4800$ \AA. North is up and East is to the left.(For a color version of the figure see the electronic version of the paper).}
\label{muse}
\end{figure}

The data reduction of the instrumental raw data from the 24 CCDs was performed by using the MUSE pipeline \citep{weilbacher}. This procedure provides a 'reduced' datacube (two spatial and one wavelength axis) where bias subtraction, flat fielding, flux and wavelength calibration are properly taken into account.  Correction for instrumental and atmospheric effects, geometrical calibration and sky subtraction are also performed within the context of this pipeline. As a result of this data reduction procedure, Fig. \ref{muse} shows the image of NGC\,330 at $\sim 4800$ \AA.

The region of NGC\,330 covered by our MUSE observations is comparable with the annulus 'A' of \cite{robertson} and allows us to analyze a sample of stars in the very center of the cluster, where the field contamination is low and most of the stars are members. On the other side this implies that the effect of crowding in NGC\,330 should be evaluated. Since we intend to derive a color-magnitude diagram of the brightest stars populating NGC\,330, 
two broadband images (using the $V$ and $I$ pass-bands as defined by \citealt{landolt92}) were extracted from the datacube. 
In this way, we were able to derive the photometry and color-magnitude diagram directly from our data. 

\begin{figure}
\center
\includegraphics[trim= .1cm 1cm 1cm 2cm, clip=true,width=.9\columnwidth]{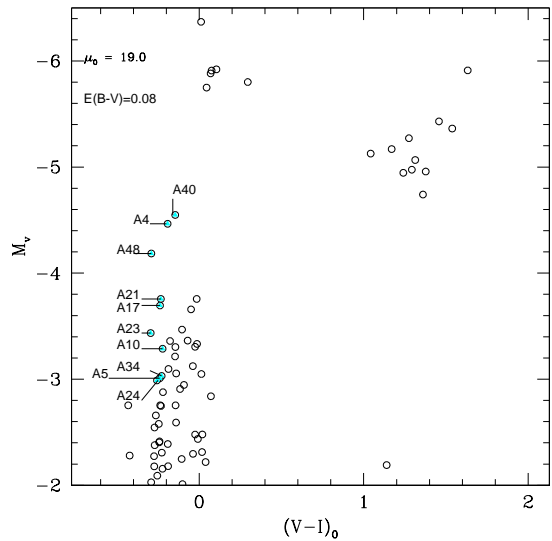}
\vspace{-1.3cm}
\caption{Observed CMD of NGC\,330. The name of the stars (designation from \protect\cite{robertson}) analyzed in this work are also reported.}  
\label{zoom}
\end{figure}

\begin{figure}
\centering
\includegraphics[trim= .1cm 0.2 0.1cm 0.2cm, clip=true,width=0.9\columnwidth]{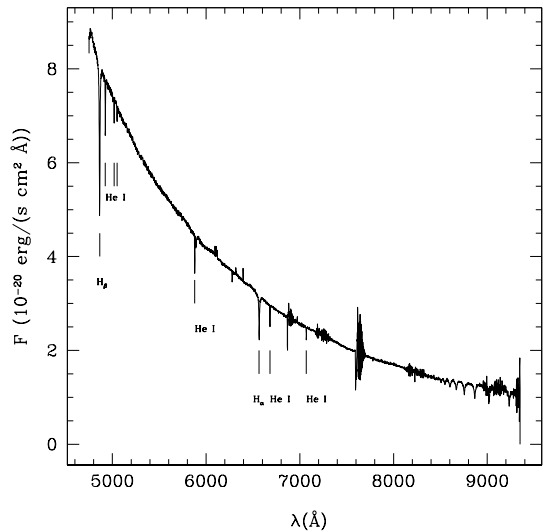}
\vspace{-2cm}
\caption{The entire spectrum of the B star A17 in NGC\,330 obtained with MUSE@VLT.}
\label{spectraA17}
\end{figure}

Moreover, we extracted from MUSE database the spectra of a sample of ten B stars (Fig. \ref{zoom}). In particular, we selected one by one the stars suffering less severe crowding and, in case of moderate overlapping point spread functions (PSFs), we took particular care in extracting the spectra from  the central region of the PSF 
 where the flux of nearby star was negligible. Clearly, this implies a slightly lower signal-to-noise ($S/N$) for the selected star but minimizes the contamination of nearby star in the extracted spectra. Taking advantage of the PSF analysis performed with \textsl{daophot} (\citealt{stetson87}), we obtained spectra which contain most of the flux of the PSF ($\geq 80\%$). Low values ($ \sim 80\%$) are due to the safe selection we adopted in avoiding the contamination from nearby stars. 
Therefore, we expect a negligible impact of crowding on the spectrum extraction.
The whole spectrum of one of the selected stars (namely, A17) is shown in Fig. \ref{spectraA17} where the typical steepness and features of B stars can be recognized.

The spectra were also corrected for atmospheric absorption features. We used the software Molecfit, a  tool distributed by ESO and based on synthetic transmission spectra calculated by a radiative transfer code \citep{molecfit1,molecfit2}.
\textsl{Molecfit} models the most appropriate atmospheric
profile (i.e., the variation in temperature, pressure, and humidity
as a function of the altitude) at the time of the given science
observations.
As input it requires \textsl{ambient parameters} (e.g. telescope altitude angle, humidity, pressure, ambient temperature, mirror temperature, elevation, longitude, latitude), \textsl{instrumental parameters}  (e.g. slit width, pixel scale), \textsl{molecular columns} (i.e. which molecules have to be considered, which depend on the spectral region analyzed), \textsl{background and continuum} (e.g. flux unit, polynomial fit for continuum), \textsl{spectroscopic resolution}, and so on.
In the wavelength range of MUSE the relevant molecules are $O_{2}$ and $H_{2}O$.
A single atmospheric profile is compiled from data from three sources: a standard atmospheric
profile for a given climate zone (produced for Michelson Interferometer for Passive Atmospheric Sounding on board the ENVISAT satellite), Global Data Assimilation System (GDAS) model profile, and the corresponding ground-based by the ESO Meteo Monitor measurements (EMM). The first one includes information on pressure, temperature, and molecular abundances  as function of height.
The second one is  provided by the National Oceanic and
Atmospheric Administration
(NOAA)\footnote{http://140.90.198.158/pub/gdas/rotating}, which gives information about the weather forecast for the location of Cerro Paranal to an altitude of $\sim$ 26 km.
The third one provides information on the local meteorological conditions in the ESO site  Paranal taken from a local meteo station mounted on a 30 m  high.
After several iterations of the $\chi^2$ minimization procedure, \textsl{Molecfit} writes the best-fit spectrum for telluric correction. This process takes into account the optimization for scaling the wavelength grid and the resolution of the model.
On the basis of this fit, the code calculates the atmospheric transmission for the wavelength range of the input spectrum and corrects it for telluric absorption. We applied this procedure for each spectrum.

\section{Photometry and comparison with isochrones}

The $V$ and $I$ images have been analyzed by using the \textsl{daophot} package developed to perform stellar photometry in crowded fields \citep{stetson87}.
The detection threshold to 5$\sigma$ above the background level was adopted. We selected a dozen of stars external to the core of the cluster to find the best PSF.
Instrumental $v$ and $i$ magnitudes were converted into the standard  V, and I  Johnson/Cousins system using, as calibrators, 12 stars in the MUSE field already detected and  with magnitudes measured  by \cite{udalski98,udalski08}\footnote{http://ogledb.astrouw.edu.pl/$\sim$ogle/photdb/index.html}.
The resulting adopted calibration equations  are:
\begin{displaymath}
V = v - [0.009 \pm 0.009](v-i) + [3.814\pm 0.008]
\end{displaymath}
\begin{displaymath}
I = i + [0.009 \pm 0.009  ](v-i) + [2.526 \pm 0.009]
\end{displaymath}

The uncertainties in the coefficients are the formal errors in the linear regression. The range  in colors of the Udalski  reference stars used as calibrators is  inside the interval  -0.14 $<$ $V-I$ $<$ 1.61. 
The color coverage of the adopted standards is is totally adequate at the blue and red extremes,  leaving slightly outside only a couple of red giants.

\begin{figure}
\center
\includegraphics[trim= .1cm 0.2 0.1cm 0.2cm, clip=true,width=0.9\columnwidth]{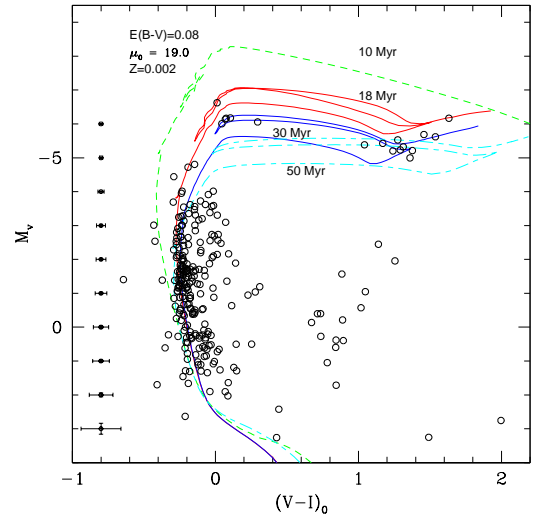}
\vspace{-2cm}
\caption{Color-magnitude diagram of the selected stars in NGC\,330.
The isochrones at 10 Myr, 18 Myr, 30 Myr and 50 Myr obtained from the PAdova and TRieste Stellar Evolution Code (PARSEC; \citealt{bressan2012, marigo2017}) are overplotted with short green dashed, red solid and blue solid and cyan short-long lines , respectively. The best-fitting isochrones are  18 Myr and 30 Myr.}
\label{iso}
\end{figure}

As a final result of the analysis performed with \textsl{daophot}, the $V$ and $I$ photometry of about 250 bright stars with a mean uncertainty of $\sim 0.01$ mag was obtained. In this sample, 34 stars are found in common with \cite{robertson} and in the following they will be indicated with the same name (i.e., with the prefix A). 
In Fig. \ref{iso} the CMD obtained with our measures is presented by assuming a distance modulus $\mu_0$ of 19.0 mag and a reddening $E(B-V) = 0.08$ mag (see \citealt{caloi}, \citealt{keller00}).
In the same figure, we also overplot the PARSEC\footnote{http://pleiadi.pd.astro.it/} \citep{bressan2012, marigo2017} isochrones with Z=0.002 \citep{spite91,reitermann90} and without rotation.

The quite broad main sequence extends up to $M_V \sim -4.5$ mag ($m_V$=14.5 mag), in agreement with \cite{chiosi95} and \cite{li2017} (see also reference therein), where the red and the blue groups are also clearly represented in our sample. These blue and red groups of stars, which are quite brighter than the main sequence termination, seem to be core He-burning stars populating the hot and cool limit of the well known loop of the effective temperature foreseen for stellar models of intermediate mass ($\approx 3-10$  $M_{\odot}$) stars. The loop extends from $(V-I)_0$ $\approx 0.0$ mag up to $\approx 1.5$ mag.
The extended main sequence turn-off has been already extensively discussed in \cite{chiosi95} and especially in \cite{li2017} and we confirm here the difficulty of interpreting this feature of the CMD of NGC\,330.
Nevertheless, we note that the two isochrones of 18 and 30 Myr  overlap fairly well the position of the main sequence stars. More interestingly, the two extended loops of the isochrones, which are due to stars experiencing the core He-burning phase, appear to reproduce quite precisely the luminosities and the colors of the blue and red groups of supergiants.
It is relevant to recall that stellar models of intermediate mass (of the adopted metallicity) predict that the evolutionary time of the He burning phases is spent part at low $T_{\rm eff}$ ($\leq 4000$ K) and part at very high  $T_{\rm eff}$ ($ \geq 12000$ K) \citep{stother92, brocato93, chiosi95,langer95,marigo2017}.

It should be noted that the blue supergiants group is composed of one very bright and hot star ($V \sim -6.5$ mag and $(V-I)_0 \sim 0.0$ mag) just overlapping the blue part of the He-burning loop of the 18 Myr old isochrone. A subgroup of four stars are $\approx 0.5$ mag fainter than this star and overlap perfectly the bluest part of the He-burning loop of the 30 Myr old isochrone.

The CMD seems to suggest that also the blue supergiants show a spread/separation in magnitude and colors. If these precise overlaps of He-burning loops are considered real features of the NGC\,330 stellar population, we find that two isochrones
separated by $\simeq 12$ Myrs are able to reproduce the data adopting reasonable distance, reddening and metallicity values. This result is compatible with the findings of \cite{li2017}.
In order to provide a quantitative example of how the isochrones change if different  ages are assumed, we plot in Fig. \ref{iso} two isochrones with 10 Myr and 50 Myr. The figure shows that the isochrone with 10 Myr  does not match the location of the observed stars, while the 50 Myr one  reproduce the MS stars but the coolest (and more populated) part of the core-he burning phase is much redder than the location of the red clump stars.
We stress here that our outcome is based on very few supergiants and needs further confirmation. Nevertheless, one single isochrone is not able to fit all the supergiants of NGC\,330 and at least two isochrones are required to fit the stars in the He-burning phase due to the observed spread/separation in their magnitudes ($\Delta V \simeq 0.5$ mag). In fact, it is quite unlikely that 
these supergiants stars  ($m_V \leq 14.0$ mag) 
have differential uncertainties of the order of $\Delta M_V \simeq 0.5$ mag, which should be needed to reconcile the observed magnitude to one single isochrone.

Finally, we have to recall that the presence of binaries can not be excluded. This occurrence could also explain the location of the brightest stars in the CMD that we address as the youngest population. We suggest here that higher resolution spectroscopy of these stars is required to provide conclusive results on this issue.

\subsection{Effect of stellar rotation}

We dedicate a paragraph to the discussion on stellar rotation, because, even if our spectra do not have the resolution high enough to analyze this issue in details, it is known that many OB stars are fast rotators ($v\sin i$ up to 300 km/s). In particular, \cite{hunter08} analyzing  high-resolution VLT-FLAMES data for  stars in  LMC and SMC, found that the mean $v \sin i$ for 77 B-type stars of NGC\,330 with masses less than 25 $M_\odot$ (as in our sample) is about 150 km/s.
Stellar rotation is a parameter directly related to the size of the convective core, (for a detailed treatment of the rotation we quote e.g. \cite{maeder00,meynet00} and references therein).
Here, we recall only the main effects of the rotation on the evolution of the B stars at low metallicity:
\begin{itemize}
    \item  the convective core of the stars increases during the MS phase;
    \item the luminosity of a rotating star is higher  (about 0.5 mag) than that  of a non-rotating  star  same mass;
    \item the combination of these two effects  implicates that the lifetime in the H-burning phase grows, but only moderately (about $20\%-30\%$ for an initial velocity of 200 km/s);
    \item the envelopes of rotating stars (angular velocity $\omega$ $>$ 0.5 $\omega_c$, with $\omega_c$ critical angular velocity) are enriched in CNO-processed material, in particular of He and N for stars with $M> 10$ $M\odot$;
    \item also during the He-burning phase the luminosity is higher because of the larger He cores, if the mass loss is not too strong;
    \item the rotation inhibits the formation of  stars in the red clump during the He-burning phase, but this  is contrasted by the mass loss effect that favor the formation of red supergiants.
    Until now the models can not reproduce the ratio between blue and red supergiants in the solar neighborhood  observed at metallicity of SMC \citep{brunish86,schaller}. Some models for the most rapid rotators at low metallicity  predict the disappearance of the red clump, and the He-burning occurs only in the blue part of the HR diagram ($\log T_{eff}$) $\sim$ 4 \citep{georgy14}.
       
\end{itemize}
We have compared our CMD, which includes also the core He-burning loop, with isochrones of rotating models from the  Geneva stellar models database\footnote{https://www.unige.ch/sciences/astro/evolution/en/database/syclist/} \citep{georgy14}, provided by the SYCLIST code. We took into account the work done by  \cite{milone18}, in which they  show  the presence of two families of stars in NGC\,330, the bluest  formed by non-rotating stars with an age compatible with 32 Myr (in agreement with our results),  the reddest  by rotating stars  with  angular velocity $\omega$ = 0.9$\omega_c$  and age around 40 Myr, representing 40-55\% of MS stars.

\begin{figure}
\centering
\includegraphics[trim= .1cm 0.2 0.1cm 0.2cm, clip=true,width=0.9\columnwidth]{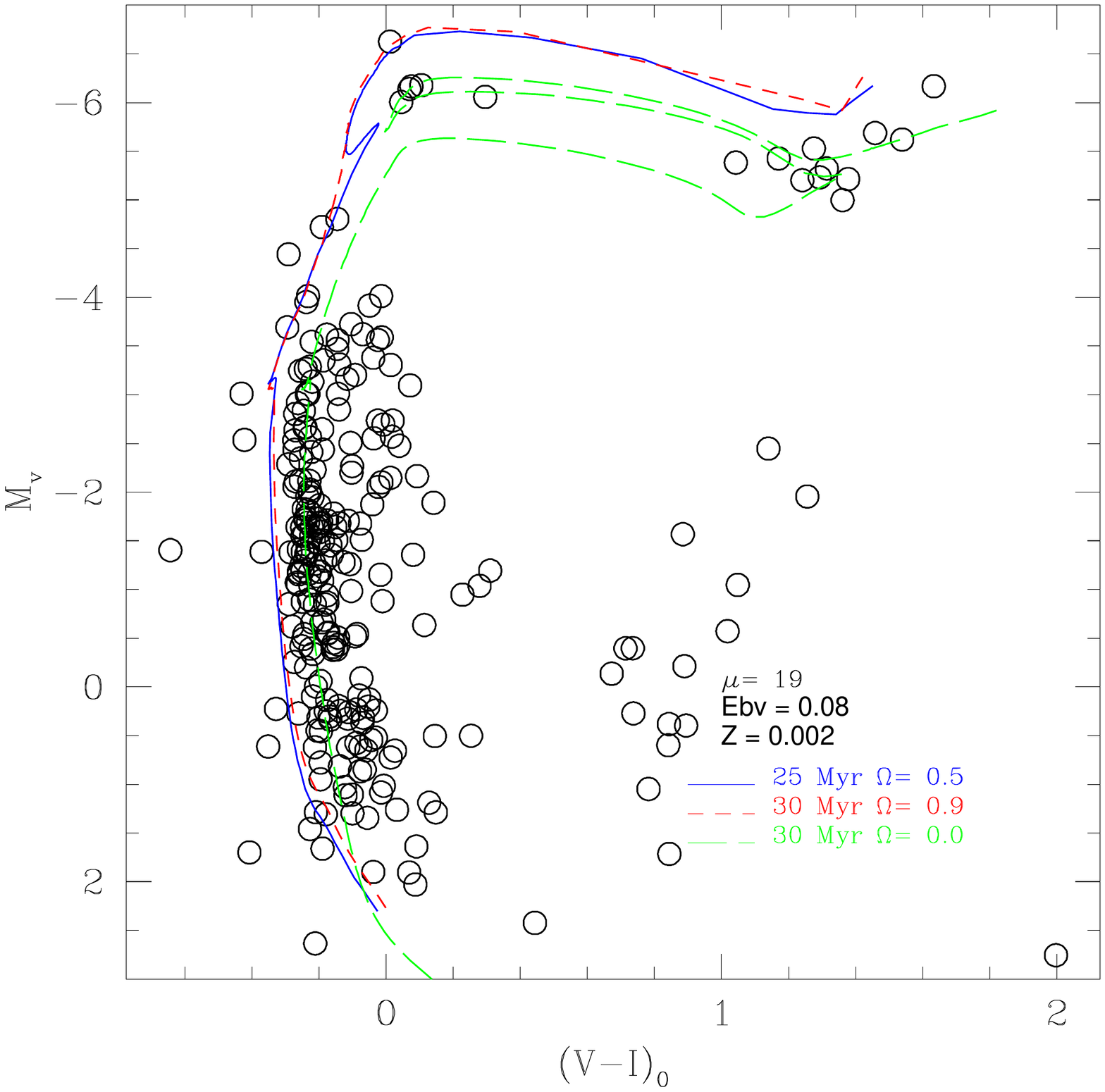}
\vspace{-2cm}
\caption{Color-magnitude diagram of NGC\,330.
The isochrone at 30 Myr  is obtained from the PAdova and TRieste Stellar Evolution Code (PARSEC; \citealt{bressan2012, marigo2017}) (the green long-dashed line);  the  isochrones computed from rotating stellar models at 25 Myr with $\omega=0.5$ $\omega_{c}$ (blue line), and  at 30 Myr with $\omega = 0.9 \omega_c$ (red dashed line) are taken from \citealt{georgy14}. The adopted values for the age, distance modulus, reddening, and metallicity are quoted in the figure.}
\label{rotation}
\end{figure}
As shown in Fig. \ref{rotation}, it is evident that the evolutionary track of Geneva predicts the absence of the red clump with this age and rotation, so they can not explain the majority of the He-burning stars. For this reason, the presence of two populations is required. One composed by non-rotating 30 Myr old stars (green long-dashed line) The second one  can be  identified  by the stars superimposed to the  isochrones of rotating models at 25 Myr with $\omega$ = 0.5 $\omega_c$(blue line) and 30 Myr with the  $\omega$ = 0.9 $\omega_c$ (red-dashed line).

In this case, we remain with a (nearly)  coeval population formed by a mixture of rotating and non-rotating stars. We keep in mind that the claim that the cluster includes a secondary rotating population is based on the location of one blue giant star, like the previous scenario where a 18 Myr population has been suggested (see Fig. \ref{iso}).
The presence of stars  with  -5.5$<$ $M_v$ $<$  -4 mag, after the overall contraction,  could favor the isochrone with $\omega$ = $0.9$ $\omega_c$ because they should be He-burning stars. 
One of these stars is  A4, already analyzed by \cite{lennon2003} who found $v sini$ = 20 km/s, thus favoring the hypothesis that the second family of stars are not fast rotators.
We remark here that the overshooting is stronger in the Padova evolutionary tracks with respect to the Geneva ones, so an error of about 5 Myr should be also taken into account.



\section{Spectra extraction and analysis}

In this section, we make use of the spectral data secured by MUSE observations to investigate the presence of a possible spread of helium abundances in the B stars of NGC\,330. Such a difference would be extremely interesting because it would support  the possible existence of two or more populations with different He content,  and would also help in alleviating the tension between models and the extended MSTO observed in NGC\,330 \citep{li2017}. Furthermore, He is a key element in the context of the multiple populations in globular clusters \citep{hb,piotto07,piotto} and discovering the presence of dichotomy or spread of He abundance within stars members of a young clusters like NGC\,330 would be extremely interesting. 
The high temperature of these B stars ensures us the presence of He {\sc i} features in their spectra. We decided to exclude Be stars to minimize uncertainties due to rotation and complexity in the spectra analysis.

\subsection{Spectroscopic Stellar Parameters}
The procedure adopted to derive the He abundances from the spectra of the selected B stars is briefly outlined in this section.

We used the \textsl{SME}\footnote{Available at http://www.stsci.edu/~valenti/sme.html}: (Spectroscopy Made Easy, version 522; \citealt{sme,sme17}) package 
to determine the fundamental parameters (effective temperature $T_{\rm eff}$,  surface gravity $\log g$, radial velocity $V_{\rm rad}$) and He abundance. 
The measurements of these quantities was performed in few key steps and synthetic spectra were computed to obtain the best-fit of the observed spectroscopic data.
\textsl{SME} is a spectral synthesis code which allows us to find the best-fit of an observed spectrum, assuming a wavelength range and initial input parameters. In particular, \textsl{SME} needs line list data for all atomic transitions of interest (i.e. element, ionization state, wavelength, excitation energy of the initial state, $\log gf$; (Vienna Atomic Line Database 3) \cite{vald}) and model atmospheres (ATLAS12, \cite{atlas12}). Plane parallel geometry, negligible magnetic field and no bulk flows are assumed by \textsl{SME}.

The parameter optimization code uses the Marquardt
algorithm \citep{marq,press92} to obtain estimates
of the parameters, through minimizing the $\chi^2$ statistic
by comparing model and observed spectra \citep{sme}.

We assumed local thermodynamical equilibrium (LTE), as the resolution of the spectra is too low to appreciate the difference in He abundance due to NLTE. The evaluation of the effect of the NLTE on our results is treated in Section. \ref{nlte}.

To derive stellar parameters removing the $\log{T_{eff}}-\log{g}$ degeneracy, we decided to fit from normalized spectra the H${\beta}$ and the HeI at 4921.9 \AA\, features of each star considering the range $4800-5000\,\AA$. 

 Before computing the synthetic spectra, input values of the stellar atmosphere have to be provided. To this purpose, we used both the  photometric information obtained from our best isochrones fit and from the direct comparison with the colors of the \cite{castelli} models. The input values for each stars of our sample can be found in Table \ref{temp} ($T_{\rm eff}^{phot}$ and $\log g^{phot}$).  Moreover, we fixed the microturbulence at 5 km/s, which is a reliable value for B stars (see \citealt{lennon2003}), and $Z=0.002$ (i.e. [Fe/H]$\sim -1$; see \citealt{spite91,reitermann90,lennon1996,lennon2003,hill}).
 The spectral resolving power ranges from $\sim $1800 at 4800 \AA \, to $\sim$ 3600 at  9300 \AA.
 


 With these inputs, we determined $T_{eff}$, $\log g$ and $V_{rad}$  from the best-fit of the spectra in the region  H${\beta}$-HeI.

\begin{figure}
\center
\includegraphics[trim= .1cm 0.2 0.1cm 0.2cm, clip=true,width=.9\columnwidth]{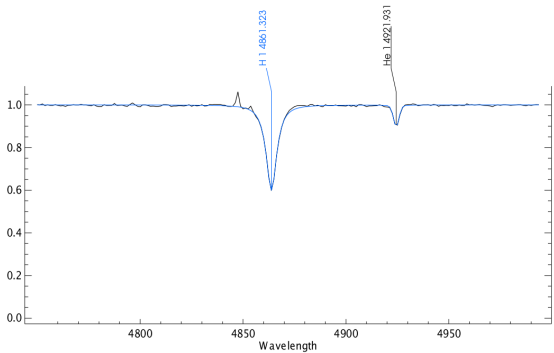}
\vspace{-.2cm}
\caption{Best-fit of the H${\beta}$ and HeI lines (labelled)  for the star A4.}  
\label{A4}
\end{figure}

An example of the best-fit of the H${\beta}$ add HeI lines is shown in Fig.\,\ref{A4} for the star A4. 

The resulting stellar parameters ($T_{\rm eff}$, $\log g$, $V_{\rm rad}$) with their uncertainties,  and the typical $S/N$ at $\lambda \sim 6000$\,\AA, are reported in Table\,\ref{temp}. We note that the reported values  of the uncertainties obtained in the  \textsl{SME} context, according with \cite{press02}, refer to intrinsic errors, and they  could  be underestimated due to systematics, like reddening, theoretical assumption, etc.

\begin{table*}
\tiny
\caption{\small{Input (photometric) and output (spectroscopic) parameters obtained from our analysis. Errors in spectroscopic parameters are listed for each target. Typical $S/N$ at about 6000\,\AA\,are reported in the last column.}}
\label{temp}
\begin{tabular}{cccccccccccc}
\hline
Star& $T_{\rm eff}^{\rm phot}$(K) & $\sigma(T_{\rm eff}^{\rm phot})$(K)& $\log{g}^{\rm phot}$&$\sigma(\log{g}^{\rm phot})$&$T_{\rm eff}^{\rm spec}$ (K) & $\sigma(T_{\rm eff}^{\rm spec})$(K) & $\log{g}^{\rm spec}$ & $\sigma(\log{g}^{\rm spec})$ &$V_{\rm rad}$(km/s)& $\sigma(V_{\rm rad})$ (km/s)  & $S/N$\\
\hline
A4&20000 &1000 & 3.0 & 0.5 & 17000 &   1000  &  2.7   &  0.1 & 152& 8 &300 \\
A5&24000  & 1000& 4.0&0.5&  22500 &  2000  & 3.7  &  0.2 & 155 & 40 &100\\
A10& 22000&1000& 4.0&0.5& 22500 &  1000 &   3.9 &   0.1 & 140 & 20 &  130\\ 
A17& 24000&1000& 4.0 & 0.5& 22000  & 1000 &   3.7 &  0.1 & 147 & 15 &200\\
A21&23000 &2000&  4.0&0.5&23500   & 1500 &   3.9  &  0.2 & 150 & 20 &200\\
A23&33000 &2000&  4.0&0.5&22000   & 1000 &   3.6  &   0.2 & 142 & 11 &200\\
A24&26000 &2000&  4.0&0.5& 23000  & 3000 &   4.1 &  0.4 & 140 & 25 & 100\\
A34&23000 &1000&   4.0&0.5&22500  & 2000  &  4.0 &  0.2 & 145 & 20 &120 \\ 
A40&16000  & 1000& 3.0&0.5&18500  &  500  &  3.2 &  0.1 & 145 & 10 & 300  \\  
A48& 32000 & 2000& 4.0 & 0.5& 25000 & 2500  & 3.6   & 0.5  &152 &45 & 200\\
\hline
\end{tabular}
\end{table*}

\subsection{Helium abundance}

\begin{figure}
\center
\includegraphics[trim= .1cm 0.2 0.1cm 0.2cm, clip=true,width=1\columnwidth]{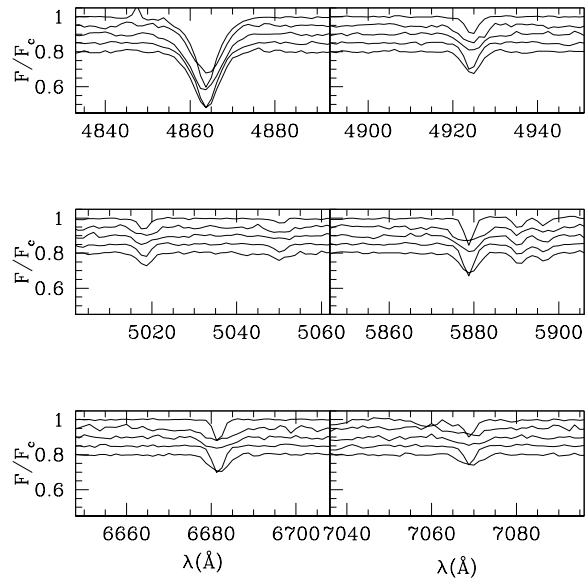}
\vspace{-2cm}
\caption{Each panel shows the region of the spectrum where the He features used in this work are located. The spectra are normalized but not red-shifted. In every panel the spectra of the stars  A4, A5, A10, A17, A21 are plotted starting from the top and shifted of 0.05 in the y-axis for graphical reasons.}
\label{all1}
\end{figure}

\begin{figure}
\center
\includegraphics[trim= .1cm 0.2 0.1cm 0.2cm, clip=true,width=1\columnwidth]{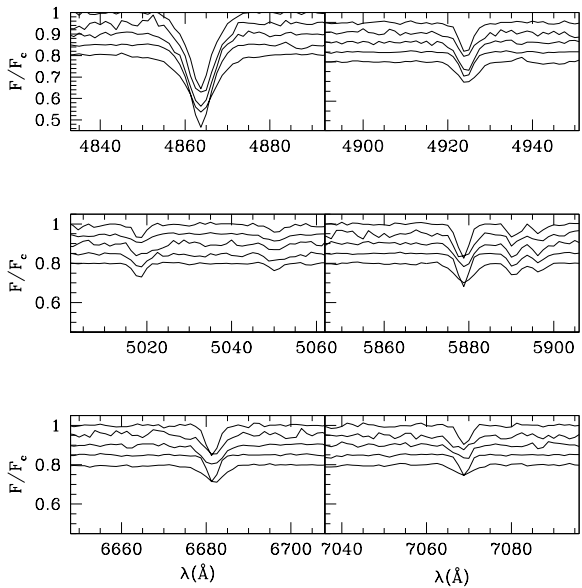}
\vspace{-2cm}
\caption{As for Fig. \ref{all1} but for the stars A23, A24, A34, A40, A48 (from top to bottom)}.
\label{all2}
\end{figure}

Once the stellar parameters of each star were evaluated, we repeated the same steps as in the previous section but for the He features listed in Table\,\ref{tablehe}, i.e we considered the following wavelength ranges: 4880-4960 \AA, 5000-5030 \AA, 5030-5100 \AA, 5860-5890 \AA, 6640-6720 \AA, 7000-7100 \AA $~$ (see Fig. \ref{all1} and Fig. \ref{all2}).
As done for the determination of the stellar parameters, also for the He features we considered the atomic and molecular line list from VALD3 \citep{vald} and the spectra were normalized to their continuum level.

Stellar parameters were constrained to the values derived by fitting the H${\beta}$ and HeI lines and then used to create the synthetic spectra around each He line. \textit{ATLAS12} model atmospheres \citep{atlas12} were considered and the best-fit was obtained around each spectral range through the \textsl{SME} package to derive the helium abundance. 
The values of the He abundance for each star and each feature are reported in Table \ref{tablehe}. Internal errors computed by \textsl{SME} and mostly due to uncertainties in atomic parameters, stellar parameters, and continuum position are also reported. Empty values in this table refer to features for which the best-fit was not obtained, in most cases because the observed line was too weak or not detectable. We also evaluated the impact of external errors considering the photometric $T_{\rm eff}$ and $log~g$. To this aim we recomputed the best-fit of each feature by adopting the photometric values of stellar parameters and derived mean errors in the final helium abundance of $\approx 0.1$ dex and $\approx 0.05$ dex due to difference choice of $T_{\rm eff}$ and $log~g$, respectively. These errors are indeed within the internal errors listed in Table\,\ref{tablehe}.
Systematic uncertainties could affect the helium abundance due to the effect of the assumption on the microturbulence. We tested the impact of  varying the microturbulence from 0 km/s to 10 km/s. The results show that an enhancement of microturbulence to 10 km/s leads to a decrease of He abundance of about 0.2 dex) with  respect the values determined with $\xi= 5$ km/s. While a value of $\xi = 0$ km/s leads to an increase of He up to 0.1 dex.
The systematic uncertainties due to the debated value of  metallicity in NGC\,330 may affect the He determination. We calculated the He abundance of the stars in our sample by assuming the lowest ([Fe/H]=-1.8, \citealt{richtler}) and the highest ([Fe/H]=-0.5, \citealt{meliani}) values commonly assumed in the literature for this cluster. It turns out that the He abundance Y in our stars changes of an amount less than 0.01 dex.

\begin{figure}
\center
\includegraphics[trim= .1cm 0.2 0.1cm 0.2cm, clip=true,width=.9\columnwidth]{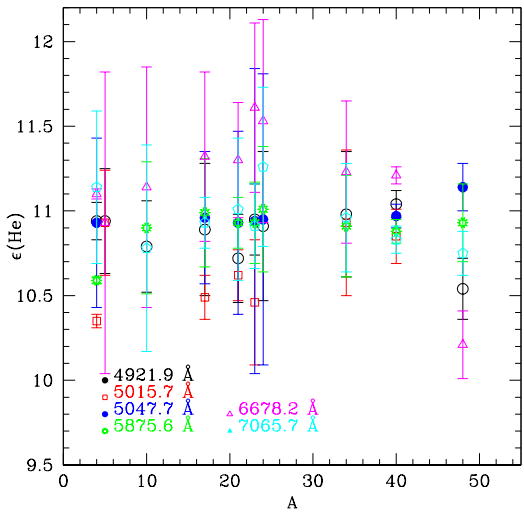}
\vspace{-2cm}
\caption{Logarithmic He abundance obtained for each helium line. Each symbol refers to different line. Uncertainties obtained by \textsl{SME} are also shown. $x$-axes reports the number of each star following the nomenclature by \protect\cite{robertson}.} 
\label{lamd}
\end{figure}

In Fig. \ref{lamd} we show the logarithmic helium abundance ($\epsilon(He)$\footnote{$\epsilon(He)$ is defined to be $\log(N_{He}/N_{H})+12$, where $N_{He}$ and $N_{H}$ are the number densities of elements He and H, respectively.} determined for each He {\sc i} line (4921.9 \AA, black circles; 5015.7 \AA, red squares; 5047.7 \AA, blue diamonds; 5875.6 \AA, green stars; 6678.2 \AA, magenta open triangles; 7065.7 \AA, cyan filled triangles) and for each star analyzed. Uncertainties obtained by \textsl{SME} are also displayed.  

We note that the He abundance determined for the 6678.2 \AA \,\ is systematically higher than ones obtained from the other lines, except for the stars A5 and A48. This could  be due to the contribution of the $^{3}He$ isotope enhancement \citep{schneider18} which can not be resolved in our spectra.

 We then computed the mean of the He abundance values derived from the best fitted features for each star. This computation was afterwards repeated using only the values deviating less than $1 \sigma$ from the initial average evaluation. This method allowed us to minimize the impact of single unpredictable errors provided by the best fitting procedure of each He feature. The He abundances averages and the standard deviation $\sigma$ of the average are reported in the $8_{th}$ and $9_{th}$ columns of Table \ref{tablehe}, while the last column lists their corresponding helium mass fraction value ($Y$).

In order to check the reliability of our measurements, we also evaluated the weighted mean of the He abundances, i.e. using as weight the error on the $\epsilon(He)$ value obtained by fitting each single feature.
As expected, we found that the two determinations of the He abundance are in full agreement within their internal uncertainties.

We notice that \cite{lennon2003} have obtained spectra of a sample of 7 stars in NGC\,330 and one star (A4) is in common with our work. The $T_{\rm eff}$ and the surface gravity we derive in this work are in  good agreement with the values obtained by these authors for the same star. While, the He abundance found by \cite{lennon2003} using nine lines is $\epsilon(He)= 10.66$, which is lower than the value found by us (i.e. $\epsilon(He) = 10.94 \pm 0.22$). 


\begin{table*}
\scriptsize
\caption{Helium abundance ($\epsilon(He)$) derived from our analysis for each line. Effective temperature and surface gravity are shown in Table \ref{temp}.}
\label{tablehe}
\begin{tabular}{ccccccccccc}
\hline
Star&$\epsilon(He)$&$\epsilon(He)$&$\epsilon(He)$&$\epsilon(He)$&$\epsilon(He)$&$\epsilon(He)$& $<\epsilon(He)>$ & $\sigma$ & $Y$ & $\sigma$(Y) \\ 
& ($\lambda$4921.9 \AA) & ($\lambda$5015.7 \AA)  & ($\lambda$5047.7 \AA)  & ($\lambda$5875.6 \AA) & ($\lambda$6678.2 \AA) & ($\lambda$7065.7 \AA) &  &   & \\ 
\hline
A4 & 10.94 $\pm$ 0.11 & 10.35 $\pm$ 0.05 & 10.93 $\pm$ 0.50 &10.59 $\pm$ 0.02&  11.10 $\pm$ 0.03& 11.14 $\pm$  0.45 &10.94   &    0.20   &  0.26 & 0.18 \\
A5& 10.94 $\pm$ 0.30 & 10.93 $\pm$ 0.30 & ...  & ...  &   10.93 $\pm$ 0.89 & ...   & 10.93 &   0.01  &  0.25  & 0.01\\
A10& 10.79 $\pm$ 0.27 & ...& ...  & 10.9  $\pm$ 0.39 & 11.14 $\pm$  0.71& 10.78 $\pm$ 0.61 &10.90   &    0.17  &  0.24 & 0.14\\
A17& 10.89 $\pm$ 0.39 &10.49 $\pm$ 0.13 &10.96 $\pm$ 0.39 & 10.99 $\pm$ 0.32 & 11.32 $\pm$ 0.50 &10.93 $\pm$ 0.15& 10.94   &    0.04   &  0.26 & 0.04\\
A21  & 10.72 $\pm$ 0.26& 10.62 $\pm$0.15 &  10.93 $\pm$ 0.54  &  10.93 $\pm$ 0.15 & 11.30 $\pm$ 0.34& 11.01 $\pm$ 0.42 &10.90   &    0.12  &  0.24 & 0.10\\  
A23&  10.95 $\pm$ 0.21 & 10.46 $\pm$ 0.37& 10.94 $\pm$ 0.90 & 10.93 $\pm$ 0.24 & 11.61 $\pm$ 0.5 & 10.91 $\pm$ 0.25& 10.93  & 0.02  &  0.25 & 0.02\\ 
A24& 10.91 $\pm$ 0.44& ...  &  10.95 $\pm$ 0.86&  11.01 $\pm$ 0.37 &  11.53 $\pm$ 0.60 & 11.26 $\pm$ 0.47& 11.03   &    0.16 &    0.30 & 0.15\\
A34& 10.98 $\pm$ 0.37 & 10.93 $\pm$  0.43  & ... & 10.91 $\pm$  0.30 &  11.23 $\pm$ 0.42& 10.96 $\pm$ 0.32& 10.95  &     0.03   &  0.26& 0.03\\
A40&  11.04 $\pm$ 0.08& 10.85 $\pm$ 0.16 & 10.97 $\pm$ 0.07 & 10.88 $\pm$  0.07& 11.21 $\pm$ 0.05& 10.83 $\pm$ 0.08 & 10.91 &   0.09  &   0.25& 0.07\\ 
A48&  10.54 $\pm$ 0.18 &  ...  &  11.14 $\pm$ 0.14  &  10.93 $\pm$ 0.23 &  10.21 $\pm$ 0.20 &  10.75$\pm$    0.13& 10.84   &    0.20   &  0.22 & 0.15\\
\hline
\end{tabular}
\end{table*}

In the following, we report the results of helium abundance for  each B star analyzed in the present work.

\subsubsection*{A4}
The spectrum of the star A4 presents one of the highest $S/N$ ratio of our sample, with a value of about $300$. The spectral regions where the He features used in this work are located are shown in Fig. \ref{all1} (first solid lines from the top). 
The line 5015.7 \AA $~$ is faint but the shape is regular and the best-fit was obtained, even if the resulting He abundance for this feature is very low ($\epsilon(He)$=10.35).
As already mentioned, this is the only star of our sample for which the He abundance has been reported by other authors \citep{lennon2003}. They derived for this target spectroscopic values of $T_{\rm eff}$= 18000 K, $\log g$ =2.8 $\pm$ 0.2 and microturbulence $\xi$= 5 km/s. These quantities are in very good agreement with our results ($T_{\rm eff}$ $\sim$ 17000 $\pm$ 1000\,K, $\log g$ =2.7 $\pm$ 0.1) and with our choice to fix the microturbulence at the value of 5 km/s (see also Table \ref{tablehe}).

However, we note that our  mean value of the He abundance ($\epsilon(He)$ = 10.94 $\pm$ 0.22(1$\sigma$)) is higher (more than $1\sigma$) than the one found by \cite{lennon2003} ($\epsilon(He)$)=10.66.  We remark here that the abundance of this star is one of the lowest of the entire sample observed by \cite{lennon2003}. 

\subsubsection*{A5}
 $S/N$ ratio of this star is about 100. The features at $\lambda=$5047.7 \AA\,, $\lambda=5875.6$ \AA\,, and $\lambda=$7065.7 \AA\, are noisy and hardly detectable, therefore no best-fit was obtained. This star shows features larger than most of the other stars, thus suggesting very high rotation (see, in particular, the features centered at 5875.6 \AA\, and 6678.2 \AA).
 The mean value of the He abundance is $\epsilon(He)$ =10.93 $\pm$ 0.01.

\subsubsection*{A10}
The effective temperature and the surface gravity derived from the synthetic spectra are in very good agreement with the photometric values (see Table\,\ref{tablehe}). For the computation of the He abundance, we used all lines with the exception of the features at $\lambda=$5015.7 \AA\, and $\lambda=$5047.7 \AA. As found for A5, also this star shows very wide features, suggesting high rotation (see solid lines in the middle of each panel of Fig.\,\ref{all1}).
In this case the derived He abundance resulted to be slightly sub-solar (i.e. 10.90$\pm$0.17), if we assume $\epsilon(He)_\odot=10.93$ as helium solar abundance (\citealt{asplund2009}).

\subsubsection*{A17}
The effective temperature and the surface gravity are in agreement, within the uncertainties, with the theoretical values and all the selected He features were used to derive the final He abundance, which resulted to be  $\epsilon(He)$= 10.94 $\pm$ 0.04

\subsubsection*{A21}
Again the effective temperature and $\log g$ are in very good agreement with the photometric values. 
The lines are particularly broadened for this star, and the uncertainties in the determination of $\epsilon(He)$ is between 0.5 and 0.6 dex. The mean value is $\epsilon(He)$=10.90 $\pm$ 0.12. 

\subsubsection*{A23}
From the photometry, we found that A23 is the target of our sample with the highest effective temperature (i.e. 33000 K), while the result of the spectral best fitting of the H${\beta}$  and HeI lines is very different (i.e. 22000 K). Also in this case, all the He features were used for the calculation of the mean He abundance, resulting to be solar.

\subsubsection*{A24}

Spectroscopic effective temperature and surface gravity are in agreement, within the errors, with the theoretical values, despite the $S/N$ ratio is one of the lowest within our sample ($\approx 100$). 
Almost each line could be considered to derive He abundance, with exception of $\lambda$5015.7 \AA, with errors up to 0.4-0.5\,dex. Accordingly, relative errors on effective temperatures are the largest one. The weighted average He abundance is $11.03 \pm 0.16$, the highest value of our sample.  As for A5, the features are broadened, pointing out a possible high rotational velocity. 

\subsubsection*{A34}
Also this star shows a relatively low $S/N$ ratio ($\approx 120$) and consequently 
this leads to a poor determination of helium abundances from the single features, with uncertainties up to $\sim 0.5$ dex. The final value of $<\epsilon(He)>$ is indeed close to the solar value ($\epsilon(He)$ = 10.95 $\pm$ 0.03; see Table\,\ref{tablehe}).

\subsubsection*{A40}

The $S/N$ ratio of this star is very high, being larger than 350.
The effective temperature from the spectra is slightly higher than the theoretical one, while the $\log g$ is in good agreement. 
All the He features were used to derive the He abundance; 
the He mean value is, $\epsilon(He)$ = 10.91 $\pm$ 0.09.
  
\subsubsection*{A48}
Our spectroscopic best-fit confirms the theoretical value of  $ \log g$ within the uncertainties, but not the $T_{eff}$, that is lower of 7000 K.
The lines of this target seem to suggest high $v \sin i$ (see Fig.\ref{all2}).
We fitted each line except for $\lambda$5015.7 \AA\,, $\epsilon(He)$=10.84 $\pm$ 0.22.

\subsection{Stellar Rotation}

As already discussed in the previous sections, the stars in our sample could rotate and this may affect the spectral features. Unfortunately, the resolution of our spectra is too low to evaluate  stellar rotation velocities  lower than 150 km/s.  We made an attempt to evaluate the contribution of the stellar rotation in the helium abundance.
To this aim, we repeated the same procedure performed in Sect. 4.1 by adopting a rotating model framework.
 We first used the isochrone of 30 Myr with the angular velocity $\omega$=$0.9\omega_c$, already shown in Fig. \ref{rotation}, to evaluate the input parameters ($T_{\rm eff}^{\rm phot}$,$\log g^{\rm phot}$, $v \sin i$) for each star. These values are then used to compute the best fit of the spectra in the spectral regions of the $H_{\beta}$ and He I at 4921.9 \AA\, lines. Once the best fit is reached, we estimated again the abundance of helium by fitting each of the available lines.
The results are reported in Tables. \ref{rot} and \ref{rothe}, where we show the photometric values considered as input parameters, the ones resulting from our best fit, and the mean abundance of He, respectively.

\begin{table*}
\scriptsize
\caption{\small{Input (photometric) and output (spectroscopic) parameters obtained from our analysis considering the stellar rotation velocity. Errors in spectroscopic parameters are listed for each target.}}
\label{rot}
\begin{tabular}{ccccccccccc}
\hline
Star & $T_{\rm eff}^{\rm phot}$(K) & $\sigma(T_{\rm eff}^{\rm phot})$(K)& $\log{g}^{\rm phot}$&$\sigma(\log{g}^{\rm phot})$&$v \sin i^{\rm input}(km/s)$ & $T_{\rm eff}^{\rm spec}$ (K) & $\sigma(T_{\rm eff}^{\rm spec})$(K) & $\log{g}^{\rm spec}$ & $\sigma(\log{g}^{\rm spec})$ & $v \sin i^{\rm output}(km/s)$ \\
\hline
A4 & 16800 & 1000 & 3.0 &0.5 & 100 & 17000 & 1000   & 2.9 & 0.1 & 100 \\
A5 & 16000 & 1000 & 3.8 & 0.5 & 350 & 21400 & 3000 & 3.5 & 0.2 & 250  \\
A10& 18500 & 1000 & 3.2 & 0.5 & 250 & 20500 & 1000 & 3.7 & 0.1 & 200  \\
A17& 20000 & 1000 & 3.3 & 0.5 & 180 & 21700 & 1000   & 3.6 & 0.1 & 100   \\
A21& 19000 & 1000 & 3.3 & 0.5& 170 & 21000 & 1000   & 3.6 & 0.2 & 150  \\
A23& 22000 & 1000 & 3.5 & 0.5 & 230 & 21600 & 2000 & 3.6 & 0.2 & 100  \\
A24& 20500 & 1000 & 3.4 & 0.5 & 330 & 21700 & 1500 & 3.9 & 0.3 & 150 \\
A34& 17800 & 1000 & 3.8 & 0.5 & 300 & 20500 & 1500 & 3.8 & 0.2 & 150 \\ 
A40& 16500 & 1000 & 2.9 & 0.5 & 100   & 17500 & 500   & 3.1 & 0.1 & 100  \\  
A48& 18000 & 1000 & 3.1 & 0.5 & 150 & 22000 & 1500 & 3.1 & 0.3 & 150 \\
\hline
\end{tabular}
\end{table*}

\begin{table*}
\begin{center}
\scriptsize
\caption{\small{Mean values, within 1$\sigma$ of the He abundance for each star of the sample obtained considering stellar rotation.}}
\label{rothe}
\begin{tabular}{ccccc}
\hline
Star &  $<\epsilon(He)>_{\rm rot}$ & $\sigma$ & $Y_{\rm rot}$ & $\sigma$ ($Y$) \\
\hline
A4  &  10.90 & 0.3  & 0.24 & 0.25\\
A5  &  11.02 & 0.03 & 0.29 & 0.03 \\
A10&  10.96 & 0.04 & 0.27 & 0.04 \\
A17&  10.98 & 0.03 & 0.27 & 0.03 \\
A21&  10.98 & 0.08 & 0.28 & 0.07 \\
A23& 10.98 & 0.10 & 0.28 & 0.10 \\
A24&  11.08 & 0.20 & 0.32 & 0.20  \\
A34&  11.11 & 0.10 & 0.34 & 0.10 \\ 
A40&  11.02 & 0.10 & 0.30 & 0.09    \\  
A48&  10.88 & 0.21 & 0.23 & 0.17  \\
\hline
\end{tabular}
\end{center}
\end{table*}

The photometric and spectroscopic values of the temperatures and gravities are in a good agreement, within the errors, but the A5 and A48.
It is interesting to note that, the obtained $T_{\rm eff}^{\rm spec}$ (K) and $\log{g}^{\rm spec}$ values are also in fair agreement, within the uncertainties, with the ones found from non-rotating models.

The fitting procedure disclosed that, for some stars as A5, A17 and A23, the projected rotation velocity obtained from spectroscopy moves away from the input values, typically decreasing their values. The highest $v sin i$ are found for the stars A5 and A10, as also evident from Fig. \ref{all1}. In fact, when the stellar rotation is taken into account, the quality of the fit  improves and the best fit is obtained for nearly all the available He lines.

\begin{figure}
\center
\includegraphics[trim= .1cm 0.2 0.1cm 0.2cm, clip=true,width=.9\columnwidth]{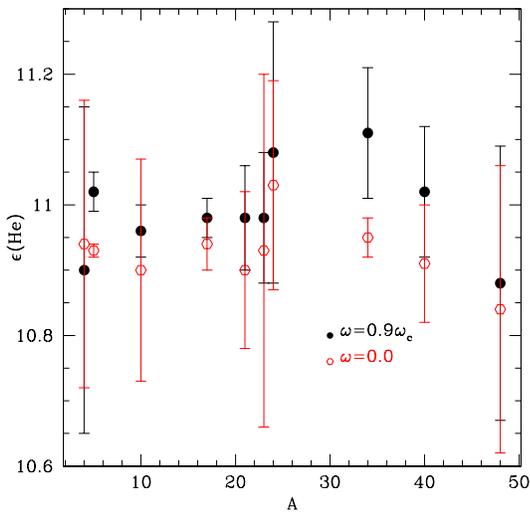}
\vspace{-1.5cm}
\caption{Mean value of the He abundance for each target considering $\omega$ = 0.9$\omega_c$ (black points), and $\omega$=0.0 (red points). Error bars are also shown.} 
\label{meanrot}
\end{figure}

Fig. \ref{meanrot} shows the helium abundance of each star. Black filled and red empty circles represent respectively the mean helium abundance considering or not the stellar rotation.
The measured mean abundances determined considering the rotation  are systematically higher than those calculated without rotation.
This result agrees with the expectations from the theory discussed in Sect. 3.1. 
We note that A4 is an exception, with the mean value obtained considering the rotation slightly lower than the one determined without rotation. For this star is not possible to fit all spectral lines considering the stellar rotation. For this reason, we suggest that this source is not a rotating star, in agreement with \cite{lennon2003}. 

Even if the resolution of the data and the uncertainties of the results require caution, we find that the rotating framework seems to suggest higher helium abundances than the values found using non-rotating models.
The mean  $<\epsilon(He)>_{\rm rot} = 11.00 \pm 0.05$ can be compared with the value obtained for the non-rotating one
$<\epsilon(He)> = 10.93 \pm 0.05$. 

 Unfortunately, we could not reach conclusive results to determine the preferred scenario about rotation, nevertheless this work provides two possible evaluations of the helium abundances for the two cases, one for fully non-rotating massive stars and one for rotating stars.

\subsection{NLTE effect}
\label{nlte}
 Conditions of non-LTE could play a role on the estimation of the He abundance because they affect both  line cores and  wings and therefore the equivalent widths of the He lines. \citep{schneider18, prz05}.
For this reason, in this section  we briefly discuss the NLTE effect.

\cite{schneider18} analyzed $He$ $I$ lines in the optical range of several B-type main sequence stars, studying also the isotopic shift of the $^4He$  due to the presence of the $^3 He$ with the NLTE approach.
From high-resolution spectra, they found  the strongest departure from LTE approach for the He I 5875.6 \AA\, line.
They  compared  the He abundance of B-type with 20000 $< T_{eff}<$ 30000 K analyzed the NLTE approach with that  found in the literature  with the LTE approach (see their Table 4). This table shows that the values determined with the LTE are roughly in agreement, within the errors, with those found with the NLTE approach, and the systematic uncertainties on $\log(n_{he})$ evaluated are reported of the order of $\pm$ $0.1$.

Recalling that the resolution of our data is about 2000, while \cite{schneider18} used spectra with a spectral resolution between 18000 and 60000, we clarify that the NLTE analysis is not possible here, and it is beyond  the limit of the feasibility with the resolution of MUSE spectra.
Moreover, the uncertainties in the He abundances derived in this work are larger than the indetermination due to the LTE/NLTE differences evaluated by \cite{schneider18}. Thus, the NLTE approach is not adopted in our analysis.

Finally, the diffusive effect process is not taken into account in our analysis on the He abundance because  the hot massive stars are characterized by a strong mass loss and rotational mixing, which tend to limit or completely erase the effect of diffusion \citep{salaris}.

\section{Discussion}
\subsection{Cluster age}
The photometric results of the bright region of the CMD of NGC\,330 (see Section $3$) suggest once more that the blue and red groups of supergiants, clearly identified since early studies, are due to core He-burning stars (e.g., \citealt{stother92}, \citealt{brocato93}). In this work, we also suggest that at least two isochrones are needed to reproduce the position of these stars in the CMD. In the non-rotation case, the separation in age we find between the two isochrones is $\approx 12$ Myr. This  partially supports the result obtained by \cite{li2017} for four clusters in the Magellanic Clouds (including NGC\,330). In fact, Li and collaborators have found that a spread of 35-50 Myr has to be considered to explain the extended main-sequence turnoffs in these clusters, while the rotation alone could not account for this observational evidence.
On the other hand, \cite{boden} appears to support a smaller spread of the order of 5 Myr.

We hence support the suggestion by \cite{li2017} using other observables, i.e. the core He-burning stars. We point out that the best-fit of the two isochrones (at 18 and 30 Myr) were found on the basis of $\textit{differencial}$ magnitudes of very bright and well measured stars. This secures that we are dealing with differential magnitude uncertainties which are very small ($\approx 0.02$ mag). 

Uncertainties on absolute calibrations related to distance modulus, reddening, calibrations and theoretical assumptions are not expected to affect 
severely the difference in age between the isochrones fitting the data, while they may change the absolute age evaluations.

A weakness of this argument is the fact that blue supergiants in the observed CMD are a few and far from any robust statistical sample, in comparison to the large number of main sequence stars. This is consequence of the evolutionary timescales of the core He-burning, which are a factor of $\approx 10$ times shorter than the ones of the core H-burning evolutionary phase lifetimes. Nevertheless, NGC\,330 is one of the youngest and most massive star cluster in the local Universe and, for this reason, it remains a fundamental ensemble of stars to be considered in studying young stellar populations.

It is interesting to highlight here that the mass of the stars in the core He-burning phase obtained from the data of the two best-fit isochrones are $9.2 \pm 0.1$ $M_{\odot}$ and $12.5 \pm 0.2$  $M_{\odot}$ \citep{bressan2012,marigo2017}.  Since the ratio of their evolutionary time during this phase is  
$\approx 6$ (estimated form the tracks consistent with the isochrones; \citealt{bressan2012}), this is in fair agreement with the small number of stars observed at younger age ($4:1$ in our - homogeneous but not complete - sample). 

Finally, we recall a further point related to stellar rotation. As we show in Fig. \ref{rotation}, isochrones computed with stellar models which take into account rotation affect the determination of the age of the stellar populations of NGC\,330. In fact, in Sect. 3.1 we show that the CMD can be explained by one nearly coeval  stellar populations composed by a mixture of rotating and non-rotating stars.

\subsection{He abundance and age}
The presence of an age spread in the stellar population of NGC\,330 leads us to investigate the existence of possible star-to-star difference in He abundance. Using the results obtained in Sect. 4, we now investigate possible relationships between He abundance and general properties of the cluster.

\begin{figure}
\center
\includegraphics[trim= .1cm 0.2 0.1cm 0.2cm, clip=true,width=.9\columnwidth]{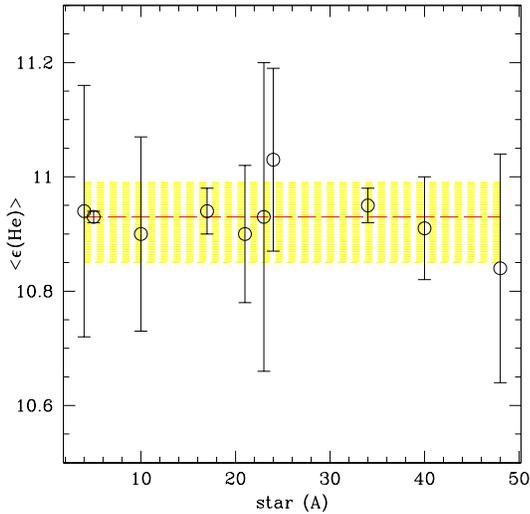}
\vspace{-1.5cm}
\caption{Mean value of the logarithmic He abundance for each target. Error bars are also shown. The red dashed line represents the average of these values, while the yellow shadow region shows the $\pm 1\sigma$ level.} 
\label{mean}
\end{figure}

We evaluated the mean He abundance of NGC\,330 as obtained considering our homogeneous set of MUSE data (we call these stars as \textsl{MUSE sample} in Table \ref{valYn330}), i.e. $\epsilon(He)_{\textsl{\small{MUSE sample}}} = 10.93 \pm 0.05$ (see also Fig. \ref{mean}).
Moreover, considering the sample of stars analyzed by \cite{lennon2003}, namely A1, A2, A4, B4, B22, B32, and B37, and the star B30  studied by \cite{korn00}, we obtain an average of $\epsilon(He)_{\textsl{other works}}=10.79\pm0.13$. Considering both He abundance results of our \textsl{\small{MUSE sample}} and of \textsl{other works}, the global value we obtain from the entire sample of stars with available He abundances is $\epsilon(He)_{\textsl{entire sample}} = 10.88 \pm 0.10$ (see Table \ref{valYn330}). In this calculation we have considered as helium abundance of the A4 star that obtained with \textsl{MUSE sample}. Therefore can conclude that the average value of the He abundance of NGC\,330 obtained with our (homogeneous) sample is consistent, within the uncertainties, with that found considering the entire - not homogeneous - sample.

\begin{table}
\begin{center}
\caption{Mean Helium abundance of NGC\,330 (see text).} 
\label{valYn330}
\begin{tabular}{ccccc}
\hline
    & $<\epsilon(He)>$& $\sigma_{<{\rm \epsilon(He)}>}$ &  $<Y>$  &  $\sigma_{<Y>}$  \\
\hline
\textsl{MUSE sample} & 10.93 & 0.05 & 0.25 & 0.02  \\
\textsl{other works} & 10.79 & 0.13 & 0.20 & 0.05  \\
\textsl{entire sample} & 10.88 & 0.10 & 0.23 & 0.04  \\
$\sim 18$ Myr stars & 10.84 & 0.13 & 0.22 & 0.05  \\
$\sim 30$ Myr stars & 10.91 & 0.08 & 0.25 & 0.03  \\
\\
\hline
\end{tabular}
\end{center}
\end{table}

We now investigate possible relationships between He abundance and stellar age. To this purpose, we tried to assign to each star of the \textsl{\small{MUSE sample}} the age of the nearest isochrone according to their CMD location. Unfortunately, the differences in color between the isochrones are too small to make a safe separation among stars located near the blue (older) or the red (younger) isochrones (see Fig. \ref{iso}). We therefore considered the $\log T_{\rm eff}-\log g$ diagram for both the \textsl{MUSE sample} and the stars analyzed in \textsl{other works}..

\begin{figure}
\center
\includegraphics[trim= .1cm 0.2 0.1cm 0.2cm, clip=true,width=.9\columnwidth]{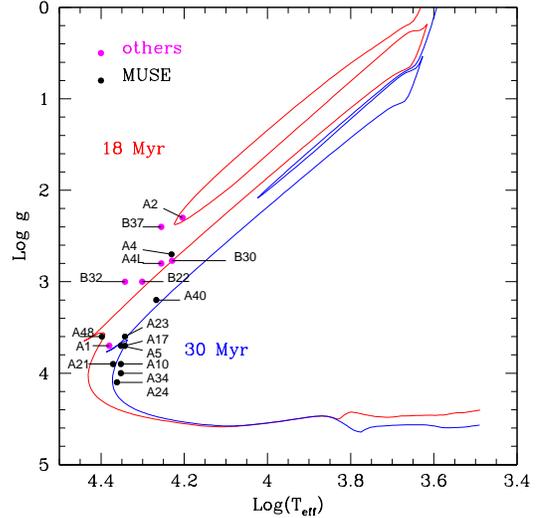}
\vspace{-1.5cm}
\caption{$\log T_{\rm eff}-\log g$ diagram for the stars analyzed spectroscopically by us with MUSE (black dots) and by other authors (magenta dots). The $Z=0.002$ PARSEC isochrones of 18 and 30 Myr are also reported with red and blue solid lines, respectively. We report also the name of the stars. We show the $\log T_{eff}$ and $\log g$ of A4 determined by  us (A4) and by \protect\cite{lennon2003}(A4L)}.  
\label{teff-grav}
\end{figure}

The $\log T_{\rm eff}-\log g$ diagram of the entire sample of NGC\,330 stars with He abundance determination is presented in Fig. \ref{teff-grav}. 

Firstly, this plot confirms our previous findings that the stars of NGC\,330 seem to show age differences/spread of $\sim$12 Myr. Then, if we arbitrarily assume from the $\log T_{\rm eff}-\log g$ diagram an age of 18 Myr for the stars 
A2, A4, A48, B22, B30, B32, and B37, and an age of 30 Myr for the stars 
A1, A5, A10, A17, A21, A23, A24, A34, and A40, the mean He abundance for each group of stars is $<\epsilon(He)>_{{\rm 18 Myr}} = 10.85\pm0.13$ and $<\epsilon(He)>_{{\rm 30 Myr}} \approx 10.92\pm0.07$, respectively (see Table\,\ref{tablehe}).
We hence find that, within the errors, the resulting He abundances of the two groups of stars separated in age do not show relevant differences.

\begin{figure}
\center
\includegraphics[trim= .1cm 0.2 0.1cm 0.2cm, clip=true,width=.9\columnwidth]{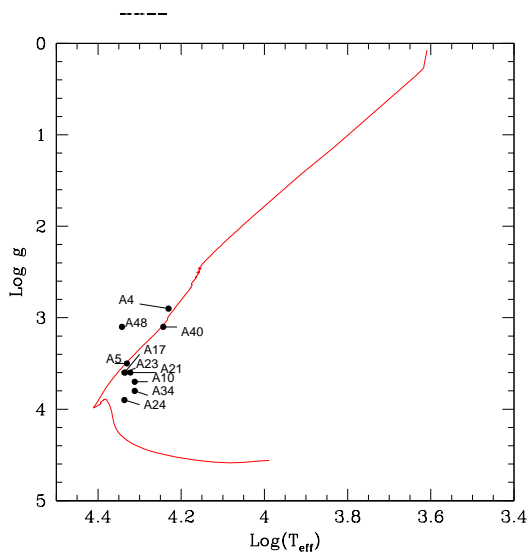}
\vspace{-1.5cm}
\caption{$\log T_{\rm eff}-\log g$ diagram for the stars analyzed spectroscopically by us with MUSE (black dots) considering stellar rotation. The $Z=0.002$ SYCLIST isochrones of  30 Myr and $\omega$ = $0.9 \omega_c$ (\citealt{georgy14}) is reported with red  solid line.  We report also the name of the stars. }.  
\label{tgrot}
\end{figure}

If the rotational scenario is taken into account, the $\log T_{\rm eff}-\log g$ diagram changes (see Fig. \ref{tgrot}).
Firstly because the fits of the spectra provide different values of $T_{\rm eff}$ and $\log g$ when the rotation of the star is considered, secondly the isochrone obtained by rotating stellar model has a different pattern from a "non-rotating" isochrone.

The  "rotating" isochrone of 30 Myr which reproduces the CMD of NGC\,330 in Fig. \ref{rotation} fails to match the position in the $\log T_{\rm eff}-\log g$ diagram of several stars (A24, A34, A10, A21, A23, A17, A48). On the contrary, the bright stars (A4, A5, A40) appears to be located along the pattern of this isochrone. Thus, this comparison suggests once more that a scenario in which the massive stars in NGC\,330 could have the same age but experiencing a different rotation velocity cannot be excluded.
In this plot we have not considered the sources analyzed by \cite{lennon2003} and \cite{korn00} because the authors considered their stars as non-rotating.

If the hypothesis of coeval populations is considered, we find that the mean helium abundance   within  $1\sigma$ of the only  "rotating" stellar population is $\sim$ 10.98 ( Y $\sim$ 0.28), slightly  higher than the mean value obtained for the non-rotating stars ($\sim$ 10.93, Y $\sim$ 0.25).
As already mentioned in Sect. 2, the region covered by our MUSE observations is comparable with the annulus 'A' of \cite{robertson} and our targets are located within the effective radius of the cluster ($R_{\rm eff} = 20''$; \citealt{zwart10}). We can therefore investigate in this internal cluster region the behavior of the stellar He abundance with the distance from the center.

\begin{figure}[!t]
\includegraphics[trim= .1cm 0.2 0.1cm 0.2cm, clip=true,width=.9\columnwidth]{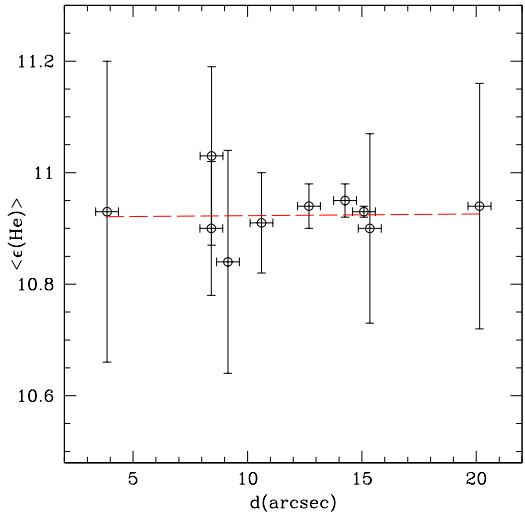}
\vspace{-1.5cm}
\caption{He abundance versus distance from the center of NGC\,330 for our MUSE sample. Dashed line represents the linear best-fit to the data. The effective radius of NGC\,330 is at $20\arcsec$. }
\label{HEvsdistance}
\end{figure}

In Fig.\,\ref{HEvsdistance}, we present our $\epsilon(He)$ values obtained using homogeneous MUSE data as a function of the distance of each star from the cluster center ($d$). The center of the cluster was defined according to the definition of \cite{evans06}, i.e. $\alpha (J2000) = 00^{\rm h} 56^{\rm m} 18.8^{\rm s}$ and $\delta (J2000) = -72^{\circ} 56{\arcmin} 18.8{\arcsec}$.
The result of our investigation clearly points out that there is no correlation between the distance of the stars from the center of the cluster and their He abundances.
This can be also quantified  by a  Kendalls' rank correlation.
For this test we found a significance of about 0.71  for the $\epsilon(He)$-distance   and  0.70 for Y-distance correlations. This means that there is no apparent correlation between the two samples of data (see Fig. \ref{HEvsdistance}).

\section{Conclusion}
We have presented an homogeneous analysis of photometric and spectroscopic data of the SMC young cluster NGC\,330. The results can be summarized as follows:
\begin{itemize}
\item We have found a possible difference in the age of the stars within NGC\,330 with a spread/separation of the order of $\approx 12 Myr$. This evidence has been derived also on the basis of the photometry of core He-burning stars.
However, 
further observational studies are required to investigate if the brightest (youngest?) stars are binaries or not.

\item  We can not exclude that the age spread is reduced or disappears, if stellar rotation is considered.
\item We have measured for the first time 
the He abundance of 10 stars placed in the center of NGC\,330 (e.g., $r_{\rm star} \leq R_{\rm eff}$). We have found a mean value of $<\epsilon(He)>=10.93\pm$0.05 for our targets homogeneously analyzed with MUSE@VLT. 
\item Considering also the stars studied in the past in this cluster, we have found a mean global helium abundance of $<\epsilon(He)>=10.88\pm$0.10, consistent with our value homogeneously derived. 
\item  We evaluated the effect of rotation on the He abundance by fitting   our spectra assuming the  $v \sin i$ value for which the fit is reached. In this case a mean global helium abundance of $<\epsilon(He)>_{rot}=11.00\pm$0.05 is found. 

\item Finally, for the main sequence B stars with $r_{\rm star} \leq R_{\rm eff}$, we have not found a possible correlation of the stellar helium abundance with the distance from the cluster center. 
\end{itemize}

The results reported in this work need more robust confirmation and we are working to increase the statistics and to minimize uncertainties.
In particular, we intend to obtain a larger sample of spectra of NGC330  B stars to increase the number of He abundance measurements  of the 18 Myr stellar population.
Moreover, we will also derive the He abundance for all the core He-burning stars of both populations.
Nevertheless, the results obtained in this paper show that MUSE@VLT is an extremely powerful instrument able to investigate both photometric and spectroscopic properties of stellar populations in young stellar clusters.  

\section*{Acknowledgements}

This work has made use of the VALD database, operated at Uppsala University, the Institute of Astronomy RAS in Moscow, and the University of Vienna. KB thanks the {\it Osservatorio Astronomico di Roma} for the hospitality during the preparation of the paper. 
We thank the anonymous referee for valuable comments and suggestions that improved the quality of the publication.
We thank Antonio Sollima for his  precious feedback.\\

\section*{Scientific software packages}

\software{daophot \citep{stetson87}, PARSEC \citep{bressan2012,marigo2017}, SYCLIST \citep{georgy14}, molecfit \citep{molecfit1, molecfit2}, SME \citep{2012ascl.soft02013V}, ATLAS12 \citep{kurucz} }



\bibliography{manuscriptbib.bib}

\begin{thebibliography}{}
\expandafter\ifx\csname natexlab\endcsname\relax\def\natexlab#1{#1}\fi
\providecommand{\url}[1]{\href{#1}{#1}}

\bibitem[{{Arp}(1959)}]{arp}
{Arp}, B.~H. 1959, \aj, 64, 254

\bibitem[{{Asplund} {et~al.}(2009){Asplund}, {Gravesse}, {Sauval}, \&
  {Scott}}]{asplund2009}
{Asplund}, M., {Gravesse}, N., {Sauval}, A.~J., \& {Scott}, P. 2009, \araa, 47,
  481

\bibitem[{{Bacon} {et~al.}(2010){Bacon}, {Accardo}, {Adjali}, {Anwand},
  {Bauer}, {Biswas}, {Blaizot}, {Boudon}, {Brau-Nogue}, {Brinchmann},
  {Caillier}, {Capoani}, {Carollo}, {Contini}, {Couderc}, {Daguis{\'e}},
  {Deiries}, {Delabre}, {Dreizler}, {Dubois}, {Dupieux}, {Dupuy}, {Emsellem},
  {Fechner}, {Fleischmann}, {Fran{\c c}ois}, {Gallou}, {Gharsa}, {Glindemann},
  {Gojak}, {Guiderdoni}, {Hansali}, {Hahn}, {Jarno}, {Kelz}, {Koehler},
  {Kosmalski}, {Laurent}, {Le Floch}, {Lilly}, {Lizon}, {Loupias}, {Manescau},
  {Monstein}, {Nicklas}, {Olaya}, {Pares}, {Pasquini}, {P{\'e}contal-Rousset},
  {Pell{\'o}}, {Petit}, {Popow}, {Reiss}, {Remillieux}, {Renault}, {Roth},
  {Rupprecht}, {Serre}, {Schaye}, {Soucail}, {Steinmetz}, {Streicher}, {Stuik},
  {Valentin}, {Vernet}, {Weilbacher}, {Wisotzki}, \& {Yerle}}]{bacon}
{Bacon}, R., {Accardo}, M., {Adjali}, L., {et~al.} 2010, in \procspie, Vol.
  7735, Ground-based and Airborne Instrumentation for Astronomy III, 773508

\bibitem[{{Bastian} \& {de Mink}(2009)}]{bastian09}
{Bastian}, N., \& {de Mink}, S.~E. 2009, \mnras, 398, L11

\bibitem[{{Bastian} {et~al.}(2013){Bastian}, {Lamers}, {de Mink}, {Longmore},
  {Goodwin}, \& {Gieles}}]{bastian2}
{Bastian}, N., {Lamers}, H.~J.~G.~L.~M., {de Mink}, S.~E., {et~al.} 2013,
  \mnras, 436, 2398

\bibitem[{{Bodensteiner} {et~al.}(2019){Bodensteiner}, {Sana}, {Mahy},
  {Patrick}, {de Koter}, {de Mink}, {Evans}, {G{\"o}tberg}, {Langer}, {Lennon},
  {Schneider}, \& {Tramper}}]{boden}
{Bodensteiner}, J., {Sana}, H., {Mahy}, L., {et~al.} 2019, arXiv e-prints,
  arXiv:1911.03477

\bibitem[{{Brandt} \& {Huang}(2015)}]{brandt15}
{Brandt}, T.~D., \& {Huang}, C.~X. 2015, \apj, 807, 25

\bibitem[{{Bressan} {et~al.}(2012){Bressan}, {Marigo}, {Girardi}, {Salasnich},
  {Dal Cero}, {Rubele}, \& {Nanni}}]{bressan2012}
{Bressan}, A., {Marigo}, P., {Girardi}, L., {et~al.} 2012, \mnras, 427, 127

\bibitem[{{Brocato} \& {Castellani}(1993)}]{brocato93}
{Brocato}, E., \& {Castellani}, V. 1993, \apj, 410, 99

\bibitem[{{Brunish} {et~al.}(1986){Brunish}, {Gallagher}, \&
  {Truran}}]{brunish86}
{Brunish}, W.~M., {Gallagher}, J.~S., \& {Truran}, J.~W. 1986, \aj, 91, 598

\bibitem[{{Caloi} {et~al.}(1993){Caloi}, {Cassatella}, {Castellani}, \&
  {Walker}}]{caloi}
{Caloi}, V., {Cassatella}, A., {Castellani}, V., \& {Walker}, A. 1993, \aap,
  271, 109

\bibitem[{{Carretta}(2015)}]{carretta15}
{Carretta}, E. 2015, \apj, 810, 148

\bibitem[{{Carretta} {et~al.}(2018){Carretta}, {Bragaglia}, {Lucatello},
  {Gratton}, {D'Orazi}, \& {Sollima}}]{carretta18}
{Carretta}, E., {Bragaglia}, A., {Lucatello}, S., {et~al.} 2018, \aap, 615, A17

\bibitem[{{Carretta} {et~al.}(2009){Carretta}, {Bragaglia}, {Gratton},
  {Lucatello}, {Catanzaro}, {Leone}, {Bellazzini}, {Claudi}, {D'Orazi},
  {Momany}, {Ortolani}, {Pancino}, {Piotto}, {Recio-Blanco}, \&
  {Sabbi}}]{carretta09}
{Carretta}, E., {Bragaglia}, A., {Gratton}, R.~G., {et~al.} 2009, \aap, 505,
  117

\bibitem[{{Castelli} \& {Kurucz}(2003)}]{castelli}
{Castelli}, F., \& {Kurucz}, R.~L. 2003, in IAU Symposium, Vol. 210, Modelling
  of Stellar Atmospheres, ed. N.~{Piskunov}, W.~W. {Weiss}, \& D.~F. {Gray},
  A20

\bibitem[{{Chantereau} {et~al.}(2019){Chantereau}, {Salaris}, {Bastian}, \&
  {Martocchia}}]{chant}
{Chantereau}, W., {Salaris}, M., {Bastian}, N., \& {Martocchia}, S. 2019,
  \mnras, 484, 5236

\bibitem[{{Chiosi} {et~al.}(1995){Chiosi}, {Vallenari}, {Bressan}, {Deng}, \&
  {Ortolani}}]{chiosi95}
{Chiosi}, C., {Vallenari}, A., {Bressan}, A., {Deng}, L., \& {Ortolani}, S.
  1995, \aap, 293, 710

\bibitem[{{D'Antona} \& {Caloi}(2004)}]{hb}
{D'Antona}, F., \& {Caloi}, V. 2004, \apj, 611, 871

\bibitem[{{D'Antona} {et~al.}(2015){D'Antona}, {Di Criscienzo}, {Decressin},
  {Milone}, {Vesperini}, \& {Ventura}}]{dantona15}
{D'Antona}, F., {Di Criscienzo}, M., {Decressin}, T., {et~al.} 2015, \mnras,
  453, 2637

\bibitem[{{Decressin} {et~al.}(2007){Decressin}, {Charbonnel}, \&
  {Meynet}}]{decressin}
{Decressin}, T., {Charbonnel}, C., \& {Meynet}, G. 2007, \aap, 475, 859

\bibitem[{{D'Ercole} {et~al.}(2016){D'Ercole}, {D'Antona}, \&
  {Vesperini}}]{dercole2016}
{D'Ercole}, A., {D'Antona}, F., \& {Vesperini}, E. 2016, \mnras, 461, 4088

\bibitem[{{D'Ercole} {et~al.}(2008){D'Ercole}, {Vesperini}, {D'Antona},
  {McMillan}, \& {Recchi}}]{dercole2008}
{D'Ercole}, A., {Vesperini}, E., {D'Antona}, F., {McMillan}, S., \& {Recchi},
  S. 2008, \mnras, 391, 825

\bibitem[{{Evans} {et~al.}(2006){Evans}, {Lennon}, {Smartt}, \&
  {Trundle}}]{evans06}
{Evans}, C.~J., {Lennon}, D.~J., {Smartt}, S.~J., \& {Trundle}, C. 2006, \aap,
  456, 623

\bibitem[{{Feast}(1991)}]{feast}
{Feast}, M.~W. 1991, in IAU Symposium, Vol. 148, The Magellanic Clouds, ed.
  R.~{Haynes} \& D.~{Milne}, 1

\bibitem[{{Georgy} {et~al.}(2014){Georgy}, {Granada}, {Ekstr{\"o}m}, {Meynet},
  {Anderson}, {Wyttenbach}, {Eggenberger}, \& {Maeder}}]{georgy14}
{Georgy}, C., {Granada}, A., {Ekstr{\"o}m}, S., {et~al.} 2014, \aap, 566, A21

\bibitem[{{Gieles} {et~al.}(2018){Gieles}, {Charbonnel}, {Krause},
  {H{\'e}nault-Brunet}, {Agertz}, {Lamers}, {Bastian}, {Gualand ris}, {Zocchi},
  \& {Petts}}]{gieles18}
{Gieles}, M., {Charbonnel}, C., {Krause}, M. G.~H., {et~al.} 2018, \mnras, 478,
  2461

\bibitem[{{Girardi} {et~al.}(2009){Girardi}, {Rubele}, \& {Kerber}}]{girardi09}
{Girardi}, L., {Rubele}, S., \& {Kerber}, L. 2009, \mnras, 394, L74

\bibitem[{{Glatt} {et~al.}(2008){Glatt}, {Grebel}, \& {Sabbi}}]{glatt}
{Glatt}, K., {Grebel}, E.~K., \& {Sabbi}, E. e.~a. 2008, \aj, 136, 1703

\bibitem[{{Goudfrooij} {et~al.}(2014){Goudfrooij}, {Girardi},
  {Kozhurina-Platais}, {Kalirai}, {Platais}, {Puzia}, {Correnti}, {Bressan},
  {Chandar}, {Kerber}, {Marigo}, \& {Rubele}}]{goud14}
{Goudfrooij}, P., {Girardi}, L., {Kozhurina-Platais}, V., {et~al.} 2014, \apj,
  797, 35

\bibitem[{{Gratton} {et~al.}(2019){Gratton}, {Bragaglia}, {Carretta},
  {D'Orazi}, {Lucatello}, \& {Sollima}}]{grat2019}
{Gratton}, R., {Bragaglia}, A., {Carretta}, E., {et~al.} 2019, \aapr, 27, 8

\bibitem[{{Gratton} {et~al.}(2012){Gratton}, {Carretta}, \&
  {Bragaglia}}]{grat2012}
{Gratton}, R.~G., {Carretta}, E., \& {Bragaglia}, A. 2012, \aapr, 20, 50G

\bibitem[{{Grebel} \& {Richtler}(1992)}]{grebel92}
{Grebel}, E.~K., \& {Richtler}, T. 1992, \aap, 253, 359

\bibitem[{{Grebel} {et~al.}(1996){Grebel}, {Roberts}, \& {Brandner}}]{grebel96}
{Grebel}, E.~K., {Roberts}, W.~J., \& {Brandner}, W. 1996, \aap, 311, 470

\bibitem[{{Hill}(1999)}]{hill}
{Hill}, V. 1999, \aap, 345, 430

\bibitem[{{Hunter} {et~al.}(2011){Hunter}, {Lennon}, {Dufton}, {Trundle},
  {Sim{\'o}n-D{\'\i}az}, {Smartt}, {Ryans}, \& {Evans}}]{hunter08}
{Hunter}, I., {Lennon}, D.~J., {Dufton}, P.~L., {et~al.} 2011, \aap, 530, C1

\bibitem[{{Kausch} {et~al.}(2015){Kausch}, {Noll}, {Smette}, {Kimeswenger},
  {Barden}, {Szyszka}, {Jones}, {Sana}, {Horst}, \& {Kerber}}]{molecfit2}
{Kausch}, W., {Noll}, S., {Smette}, A., {et~al.} 2015, \aap, 576, A78

\bibitem[{{Keller} \& {Bessell}(1998)}]{keller98}
{Keller}, S.~C., \& {Bessell}, M.~S. 1998, \aap, 340, 397

\bibitem[{{Keller} {et~al.}(2000){Keller}, {Bessell}, \& {Da Costa}}]{keller00}
{Keller}, S.~C., {Bessell}, M.~S., \& {Da Costa}, G.~S. 2000, \aj, 119, 1748

\bibitem[{{Keller} {et~al.}(1999){Keller}, {Wood}, \& {Bessell}}]{keller99}
{Keller}, S.~C., {Wood}, P.~R., \& {Bessell}, M.~S. 1999, \aaps, 134, 489

\bibitem[{{Kelz} {et~al.}(2016){Kelz}, {Kamann}, {Urrutia}, {Weilbacher}, \&
  {Bacon}}]{kelz}
{Kelz}, A., {Kamann}, S., {Urrutia}, T., {Weilbacher}, P., \& {Bacon}, R. 2016,
  in Astronomical Society of the Pacific Conference Series, Vol. 507,
  Multi-Object Spectroscopy in the Next Decade: Big Questions, Large Surveys,
  and Wide Fields, ed. I.~{Skillen}, M.~{Balcells}, \& S.~{Trager}, 323

\bibitem[{{Korn} {et~al.}(2000){Korn}, {Becker}, {Gummersbach}, \&
  {Wolf}}]{korn00}
{Korn}, A.~J., {Becker}, S.~R., {Gummersbach}, C.~A., \& {Wolf}, B. 2000, \aap,
  353, 655

\bibitem[{{Kurucz}(1979)}]{kurucz}
{Kurucz}, R.~L. 1979, \apjs, 40, 1

\bibitem[{{Kurucz}(2013)}]{atlas12}
---. 2013, {ATLAS12: Opacity sampling model atmosphere program}, Astrophysics
  Source Code Library, , , ascl:1303.024

\bibitem[{{Lagioia} {et~al.}(2019{\natexlab{a}}){Lagioia}, {Milone}, {Marino},
  {Cordoni}, \& {Tailo}}]{lagioia19}
{Lagioia}, E.~P., {Milone}, A.~P., {Marino}, A.~F., {Cordoni}, G., \& {Tailo},
  M. 2019{\natexlab{a}}, arXiv e-prints, arXiv:1909.08439

\bibitem[{{Lagioia} {et~al.}(2019{\natexlab{b}}){Lagioia}, {Milone}, {Marino},
  \& {Dotter}}]{lagioia}
{Lagioia}, E.~P., {Milone}, A.~P., {Marino}, A.~F., \& {Dotter}, A.
  2019{\natexlab{b}}, \apj, 871, 140

\bibitem[{{Landolt}(1992)}]{landolt92}
{Landolt}, A.~U. 1992, \aj, 104, 340

\bibitem[{{Langer} \& {Maeder}(1995)}]{langer95}
{Langer}, N., \& {Maeder}, A. 1995, \aap, 295, 685

\bibitem[{{Lennon} {et~al.}(2003){Lennon}, {Dufton}, \& {Crowley}}]{lennon2003}
{Lennon}, D.~J., {Dufton}, P.~L., \& {Crowley}, C. 2003, \aap, 398, 455

\bibitem[{{Lennon} {et~al.}(1996){Lennon}, {Dufton}, {Mazzali}, {Pasian}, \&
  {Marconi}}]{lennon1996}
{Lennon}, D.~J., {Dufton}, P.~L., {Mazzali}, P.~A., {Pasian}, F., \& {Marconi},
  G. 1996, \aap, 314, 243

\bibitem[{{Li} {et~al.}(2017){Li}, {de Grijs}, {Deng}, \& {Milone}}]{li2017}
{Li}, C., {de Grijs}, R., {Deng}, L., \& {Milone}, A.~P. 2017, \apj, 844, 119

\bibitem[{{Mackey} \& {Broby Nielsen}(2007)}]{mackey07}
{Mackey}, A.~D., \& {Broby Nielsen}, P. 2007, \mnras, 379, 151

\bibitem[{{Mackey} \& {Gilmore}(2003)}]{mackey}
{Mackey}, A.~D., \& {Gilmore}, G.~F. 2003, \mnras, 338, 120

\bibitem[{{Maeder} {et~al.}(1999){Maeder}, {Grebel}, \&
  {Mermilliod}}]{maeder99}
{Maeder}, A., {Grebel}, E.~K., \& {Mermilliod}, J.-C. 1999, \aap, 346, 459

\bibitem[{{Maeder} \& {Meynet}(2000)}]{maeder00}
{Maeder}, A., \& {Meynet}, G. 2000, \araa, 38, 143

\bibitem[{{Marigo} {et~al.}(2017){Marigo}, {Girardi}, {Bressan}, {Rosenfield},
  {Aringer}, {Chen}, {Dussin}, {Nanni}, {Pastorelli}, {Rodrigues}, {Trabucchi},
  {Bladh}, {Dalcanton}, {Groenewegen}, {Montalb{\'a}n}, \& {Wood}}]{marigo2017}
{Marigo}, P., {Girardi}, L., {Bressan}, A., {et~al.} 2017, \apj, 835, 77

\bibitem[{{Marino} {et~al.}(2018){Marino}, {Przybilla}, {Milone}, {Da Costa},
  {D'Antona}, {Dotter}, \& {Dupree}}]{marino18}
{Marino}, A.~F., {Przybilla}, N., {Milone}, A.~P., {et~al.} 2018, \aj, 156, 116

\bibitem[{{Marino} {et~al.}(2014){Marino}, {Milone}, {Przybilla}, {Bergemann},
  {Lind}, {Asplund}, {Cassisi}, {Catelan}, {Casagrande}, {Valcarce}, {Bedin},
  {Cort{\'e}s}, {D'Antona}, {Jerjen}, {Piotto}, {Schlesinger}, {Zoccali}, \&
  {Angeloni}}]{marino2014}
{Marino}, A.~F., {Milone}, A.~P., {Przybilla}, N., {et~al.} 2014, \mnras, 437,
  1609

\bibitem[{{Martayan} {et~al.}(2007{\natexlab{a}}){Martayan}, {Floquet},
  {Hubert}, {Guti{\'e}rrez-Soto}, {Fabregat}, {Neiner}, \&
  {Mekkas}}]{martayan07b}
{Martayan}, C., {Floquet}, M., {Hubert}, A.~M., {et~al.} 2007{\natexlab{a}},
  \aap, 472, 577

\bibitem[{{Martayan} {et~al.}(2007{\natexlab{b}}){Martayan}, {Fr{\'e}mat},
  {Hubert}, {Floquet}, {Zorec}, \& {Neiner}}]{martayan07a}
{Martayan}, C., {Fr{\'e}mat}, Y., {Hubert}, A.-M., {et~al.} 2007{\natexlab{b}},
  \aap, 462, 683

\bibitem[{{Martocchia} {et~al.}(2017){Martocchia}, {Bastian}, {Usher},
  {Kozhurina-Platais}, {Niederhofer}, {Cabrera-Ziri}, {Dalessandro},
  {Hollyhead}, {Kacharov}, {Lardo}, {Larsen}, {Mucciarelli}, {Platais},
  {Salaris}, {Cordero}, {Geisler}, {Hilker}, {Li}, \& {Mackey}}]{martocchia17}
{Martocchia}, S., {Bastian}, N., {Usher}, C., {et~al.} 2017, \mnras, 468, 3150

\bibitem[{{Mazzali} {et~al.}(1996){Mazzali}, {Lennon}, {Pasian}, {Marconi},
  {Baade}, \& {Castellani}}]{mazzali96}
{Mazzali}, P.~A., {Lennon}, D.~J., {Pasian}, F., {et~al.} 1996, \aap, 316, 173

\bibitem[{{McLaughlin} \& {van der Marel}(2005)}]{mclagh}
{McLaughlin}, D.~E., \& {van der Marel}, R.~P. 2005, \apjs, 161, 304

\bibitem[{{Meliani} {et~al.}(1995){Meliani}, {Barbuy}, \& {Perrin}}]{meliani}
{Meliani}, M.~T., {Barbuy}, B., \& {Perrin}, M.-N. 1995, \aap, 300, 349

\bibitem[{{Meynet} \& {Maeder}(2000)}]{meynet00}
{Meynet}, G., \& {Maeder}, A. 2000, \aap, 361, 101

\bibitem[{{Milone} {et~al.}(2009){Milone}, {Bedin}, {Piotto}, \&
  {Anderson}}]{milone09}
{Milone}, A.~P., {Bedin}, L.~R., {Piotto}, G., \& {Anderson}, J. 2009, \aap,
  497, 755

\bibitem[{{Milone} {et~al.}(2016){Milone}, {Marino}, {D'Antona}, {Bedin}, {Da
  Costa}, {Jerjen}, \& {Mackey}}]{milone2016}
{Milone}, A.~P., {Marino}, A.~F., {D'Antona}, F., {et~al.} 2016, \mnras, 458,
  4368

\bibitem[{{Milone} {et~al.}(2015){Milone}, {Bedin}, {Piotto}, {Marino},
  {Cassisi}, {Bellini}, {Jerjen}, {Pietrinferni}, {Aparicio}, \&
  {Rich}}]{milone15}
{Milone}, A.~P., {Bedin}, L.~R., {Piotto}, G., {et~al.} 2015, \mnras, 450, 3750

\bibitem[{{Milone} {et~al.}(2017){Milone}, {Marino}, {D'Antona}, {Bedin},
  {Piotto}, {Jerjen}, {Anderson}, {Dotter}, {di Criscienzo}, \&
  {Lagioia}}]{milone17}
{Milone}, A.~P., {Marino}, A.~F., {D'Antona}, F., {et~al.} 2017, \mnras, 465,
  4363

\bibitem[{{Milone} {et~al.}(2018){Milone}, {Marino}, {Di Criscienzo},
  {D'Antona}, {Bedin}, {Da Costa}, {Piotto}, {Tailo}, {Dotter}, {Angeloni},
  {Anderson}, {Jerjen}, {Li}, {Dupree}, {Granata}, {Lagioia}, {Mackey},
  {Nardiello}, \& {Vesperini}}]{milone18}
{Milone}, A.~P., {Marino}, A.~F., {Di Criscienzo}, M., {et~al.} 2018, \mnras,
  477, 2640

\bibitem[{{Mucciarelli} {et~al.}(2014){Mucciarelli}, {Dalessandro}, {Ferraro},
  {Origlia}, \& {Lanzoni}}]{mucciarelli14}
{Mucciarelli}, A., {Dalessandro}, E., {Ferraro}, F.~R., {Origlia}, L., \&
  {Lanzoni}, B. 2014, \apjl, 793, L6

\bibitem[{{Pasquini} {et~al.}(2011){Pasquini}, {Mauas}, {Kaufl}, \&
  {Cacciari}}]{pasquini2011}
{Pasquini}, L., {Mauas}, P., {Kaufl}, H.~U., \& {Cacciari}, C. 2011, \aap, 531,
  35

\bibitem[{{Piotto}(2008)}]{piotto}
{Piotto}, G. 2008, \memsai, 79, 334

\bibitem[{{Piotto}(2010)}]{piotto2010}
---. 2010, Publication of Korean Astronomical Society, 25, 91

\bibitem[{{Piotto} {et~al.}(2007){Piotto}, {Bedin}, {Anderson}, {King},
  {Cassisi}, {Milone}, {Villanova}, {Pietrinferni}, \& {Renzini}}]{piotto07}
{Piotto}, G., {Bedin}, L.~R., {Anderson}, J., {et~al.} 2007, \apjl, 661, L53

\bibitem[{{Piskunov} \& {Valenti}(2017)}]{sme17}
{Piskunov}, N., \& {Valenti}, J.~A. 2017, \aap, 597, A16

\bibitem[{{Piskunov} {et~al.}(1995){Piskunov}, {Kupka}, {Ryabchikova}, {Weiss},
  \& {Jeffery}}]{vald}
{Piskunov}, N.~E., {Kupka}, F., {Ryabchikova}, T.~A., {Weiss}, W.~W., \&
  {Jeffery}, C.~S. 1995, \aaps, 112, 525

\bibitem[{{Portegies Zwart} {et~al.}(2010){Portegies Zwart}, {McMillan}, \&
  {Gieles}}]{zwart10}
{Portegies Zwart}, S.~F., {McMillan}, S.~L.~W., \& {Gieles}, M. 2010, \araa,
  48, 431

\bibitem[{{Press} {et~al.}(1992){Press}, {Teukolsky}, {Vetterling}, \&
  {Flannery}}]{press92}
{Press}, W.~H., {Teukolsky}, S.~A., {Vetterling}, W.~T., \& {Flannery}, B.~P.
  1992, {Numerical recipes in FORTRAN. The art of scientific computing}

\bibitem[{{Press} {et~al.}(2002){Press}, {Teukolsky}, {Vetterling}, \&
  {Flannery}}]{press02}
---. 2002, {Numerical recipes in C++ : the art of scientific computing}

\bibitem[{{Przybilla}(2005)}]{prz05}
{Przybilla}, N. 2005, \aap, 443, 293

\bibitem[{{Reitermann} {et~al.}(1990){Reitermann}, {Baschek}, {Stahl}, \&
  {Wolf}}]{reitermann90}
{Reitermann}, A., {Baschek}, B., {Stahl}, O., \& {Wolf}, B. 1990, \aap, 234,
  109

\bibitem[{{Richtler} \& {Nelles}(1983)}]{richtler}
{Richtler}, T., \& {Nelles}, B. 1983, \aap, 119, 75

\bibitem[{{Robertson}(1974)}]{robertson}
{Robertson}, J.~W. 1974, \aaps, 15, 261

\bibitem[{{Salaris} \& {Cassisi}(2017)}]{salaris}
{Salaris}, M., \& {Cassisi}, S. 2017, Royal Society Open Science, 4, 170192

\bibitem[{{Schaller} {et~al.}(1992){Schaller}, {Schaerer}, {Meynet}, \&
  {Maeder}}]{schaller}
{Schaller}, G., {Schaerer}, D., {Meynet}, G., \& {Maeder}, A. 1992, \aaps, 96,
  269

\bibitem[{{Schneider} {et~al.}(2018){Schneider}, {Irrgang}, {Heber}, {Nieva},
  \& {Przybilla}}]{schneider18}
{Schneider}, D., {Irrgang}, A., {Heber}, U., {Nieva}, M.~F., \& {Przybilla}, N.
  2018, \aap, 618, A86

\bibitem[{{Sirianni} {et~al.}(2002){Sirianni}, {Nota}, {De Marchi},
  {Leitherer}, \& {Clampin}}]{sirianni02}
{Sirianni}, M., {Nota}, A., {De Marchi}, G., {Leitherer}, C., \& {Clampin}, M.
  2002, \apj, 579, 275

\bibitem[{{Smette} {et~al.}(2015){Smette}, {Sana}, {Noll}, {Horst}, {Kausch},
  {Kimeswenger}, {Barden}, {Szyszka}, {Jones}, {Gallenne}, {Vinther},
  {Ballester}, \& {Taylor}}]{molecfit1}
{Smette}, A., {Sana}, H., {Noll}, S., {et~al.} 2015, \aap, 576, A77

\bibitem[{{Spite} {et~al.}(1991){Spite}, {Spite}, \& {Richtler}}]{spite91}
{Spite}, F., {Spite}, M., \& {Richtler}, T. 1991, \aap, 252, 557

\bibitem[{{Stetson}(1987)}]{stetson87}
{Stetson}, P.~B. 1987, \pasp, 99, 191

\bibitem[{{Stothers} \& {Chin}(1992)}]{stother92}
{Stothers}, R.~B., \& {Chin}, C.-W. 1992, \apj, 390, 136

\bibitem[{{Tanab{\'e}} {et~al.}(2013){Tanab{\'e}}, {Motohara}, {Tateuchi},
  {Matsunaga}, {Ita}, {Toshikawa}, {Konishi}, {Kato}, \& {Yoshii}}]{tanabe13}
{Tanab{\'e}}, T., {Motohara}, K., {Tateuchi}, K., {et~al.} 2013, \pasj, 65, 55

\bibitem[{{Udalski} {et~al.}(1998){Udalski}, {Szymanski}, {Kubiak},
  {Pietrzynski}, {Wozniak}, \& {Zebrun}}]{udalski98}
{Udalski}, A., {Szymanski}, M., {Kubiak}, M., {et~al.} 1998, \actaa, 48, 147

\bibitem[{{Udalski} {et~al.}(2008){Udalski}, {Soszy{\'n}ski}, {Szyma{\'n}ski},
  {Kubiak}, {Pietrzy{\'n}ski}, {Wyrzykowski}, {Szewczyk}, {Ulaczyk}, \&
  {Poleski}}]{udalski08}
{Udalski}, A., {Soszy{\'n}ski}, I., {Szyma{\'n}ski}, M.~K., {et~al.} 2008,
  \actaa, 58, 329

\bibitem[{{Valenti} \& {Piskunov}(1996)}]{sme}
{Valenti}, J.~A., \& {Piskunov}, N. 1996, \aaps, 118, 595

\bibitem[{{Valenti} \& {Piskunov}(2012)}]{2012ascl.soft02013V}
---. 2012, {SME: Spectroscopy Made Easy}, , , ascl:1202.013

\bibitem[{W.~Marquardt(1963)}]{marq}
W.~Marquardt, D. 1963, 11, 431

\bibitem[{{Weilbacher} {et~al.}(2014){Weilbacher}, {Streicher}, {Urrutia},
  {P{\'e}contal-Rousset}, {Jarno}, \& {Bacon}}]{weilbacher}
{Weilbacher}, P.~M., {Streicher}, O., {Urrutia}, T., {et~al.} 2014, in
  Astronomical Society of the Pacific Conference Series, Vol. 485, Astronomical
  Data Analysis Software and Systems XXIII, ed. N.~{Manset} \& P.~{Forshay},
  451

\end{thebibliography}




\end{document}